\documentclass[twocolumn]{aastex63}
\usepackage{graphicx}
\usepackage{wrapfig}
\usepackage{natbib}
\usepackage{color}
\usepackage{mathtools}
\usepackage{epstopdf}
\usepackage{hyperref} %----set up hyper link to citations, equations...
\hypersetup{colorlinks=true,citecolor=blue}

\def    \gcn        {\rm {GRB Coordinates Network}}
\def    \apjl  		{\rm {ApJL}}
\def    \apj  		{\rm {ApJ}}
\def    \aa  		{\rm {A\&A}}
\def    \mnras  	{\rm {MNRAS}}
\def    \araa  		{\rm {ARA\& A}}
\def    \apjl  		{\rm {ApJL}}

\def	\cm		{\,{\rm {cm}}}
\def	\K		{\,{\rm K}}
\def	\g		{\,{\rm {g}}}
\def	\mum	{\,{\mu \rm{m}}}
\def	\erg		{\,{\rm {erg}}}
\def	\ev		{\,{\rm {eV}}}
\def	\s		{\,{\rm {s}}}
\def	\H		{{\rm {H}}}

\def \bea {\begin{eqnarray}}
\def \ena {\end{eqnarray}}                  

\begin{document}
\shorttitle{GRB Afterglows and Grain Disruption}
\shortauthors{Hoang, Giang and Tram}
\title{Gamma-ray Burst Afterglows: Time-Varying Extinction, Polarization, and Colors due to Rotational Disruption of Dust Grains}

\author{Thiem Hoang}
\affiliation{Korea Astronomy and Space Science Institute, Yuseong-gu, Daejeon 34055, South Korea; \href{mailto:thiemhoang@kasi.re.kr}{thiemhoang@kasi.re.kr}}
\affiliation{Korea University of Science and Technology, 217 Gajeong-ro, Yuseong-gu, Daejeon, 34113, South Korea}

\author{Nguyen Chau Giang}
\affiliation{University of Science and Technology of Hanoi, VAST, 18 Hoang Quoc Viet, Vietnam}

\author{Le Ngoc Tram}
\affiliation{SOFIA-USRA, NASA Ames Research Center, MS 232-11, Moffett Field, 94035 CA, USA}
\affiliation{University of Science and Technology of Hanoi, VAST, 18 Hoang Quoc Viet, Vietnam}

\begin{abstract}

Prompt optical emission of gamma-ray bursts (GRBs) is known to have important effects on the surrounding environment. In this paper, we study rotational disruption and alignment of dust grains by radiative torques (RATs) induced by GRB afterglows and predict their signatures on the observational properties of GRB afterglows. We first study grain disruption using RAdiative Torque Disruption (RATD) mechanism and find that large grains (size $>0.1 \mu\rm m$) within a distance of $d< 40$ pc from the source can be disrupted into smaller grains. We then model the extinction curve of GRB afterglows and find that optical-NIR extinction is rapidly decreased, and UV extinction increases due to the conversion of large grains into smaller ones via RATD. The total-to-selective visual extinction ratio is found to decrease from the standard value of $R_{V}\sim 3.1$ to $\sim 1.5$ after disruption time $t_{\rm disr} \lesssim 10^{4}$ s. Next, we study grain alignment by RATs induced by GRB afterglows and model the wavelength-dependence polarization produced by grains aligned with magnetic fields. We find that polarization degree first increases due to enhanced alignment of small grains and then decreases when grain disruption by RATD begins. The maximum polarization wavelength $ \lambda_{\rm max}$ decreases rapidly from the standard value of $\sim 0.55 \mum$ to $\sim 0.15 \mum$ over alignment time of $t_{\rm align} \lesssim 30$ s due to enhanced alignment of small grains. Finally, we found that RATD induces a significant decrease in optical/NIR extinction, producing an optical re-brightening in the observed light curve of GRB afterglows. We show that our theoretical predictions can explain various observational properties of GRB afterglows, including steep extinction curves, time-variability of colors, and optical re-brightening of GRB afterglows.

\end{abstract}
\keywords{gamma-ray burst:general, dust, extinction}

\section{Introduction}\label{sec:intro}
Gamma-ray bursts (GRBs) are among the most luminous transient events in the Universe. GRBs are thought to originate from a highly relativistic jet powered by a central engine (black hole or a highly magnetized neutron star--magnetar). During the burst (of $\sim 10-100$ s duration for long GRBs), prompt emission from X-ray to ultraviolet (UV)-optical wavelengths is also observed. After the prompt phase, GRB afterglows are emitted due to the interaction of relativistic jets with the ambient medium, including radiative cooling of reverse shocks and then forward shocks (\citealt{Meszaros97}). GRB afterglows can last up to days and thus offer an essential window to study the local environments around GRBs, which are required to understand the progenitors and emission mechanism of GRBs. 

The effects of dust extinction are particularly important for understanding the nature and progenitor of GRBs because GRBs are expected to occur in star-forming dusty regions (\citealt{Paczy98}). Indeed, only about $60\%$ of {\it Swift} GRBs are detected in optical wavelengths, whereas X-ray detection of GRBs is more than $95\%$ (\citealt{Gehrels:2009}). This leaves about $40\%$ of optical GRBs undetectable, so-called "dark" GRBs. The leading reason for that lies in the attenuation of optical photons by intervening dust (see \citealt{Drain02} and reference therein). 

GRB afterglows also offer a unique probe to study gas and dust properties in the interstellar medium (ISM) of high-redshift galaxies (i.e., $z>2$) due to their stable, highest intrinsic luminosity (see \citealt{Schady17} and reference therein). Observations show that the wavelength-dependent extinction (extinction curve) toward individual GRBs is described by a Small Magellanic Cloud (SMC)-like with a steep far-UV rise, which suggest predominance of small grains in the local environment (e.g., \citealt{Schady12}; \citealt{Heintz:2017}; \citealt{Zafar:2018}). The question how small grains are predominant in the local environment of GRBs is still unknown. Similar to type Ia supernovae (SNe Ia), we expect intense radiation from GRBs would have important effect on surrounding dust (\citealt{Hoang19}).

The effect of prompt optical-UV emission from GRBs on surrounding dust was first studied by \cite{Waxman20}, where the authors found that dust grains within 10 pc can be sublimated within 10 s from the burst. Later, \cite{Fruchter01} studied dust destruction by grain heating and charging (i.e., Coulomb explosions) due to X-rays, where the latter mechanism is found to be more efficient. However, the effect of grain-size dependence of photoelectric yield by X-rays (\citealt{Weingartner:2006}; \citealt{Hoang15b}) is not considered in \cite{Fruchter01}. Detailed modeling of the time-dependent dust extinction due to the thermal sublimation and ion-field emission by the optical-UV flash (i.e., prompt emission) was presented in \cite{Perna03}, where the authors found that dust extinction decreases significantly by $t\sim$ 10 s from the start of the burst.

Early-time observations of GRB afterglows (e.g., GRB 111209A by \cite{Stratta13}, GRB 120119A by \cite{Morgan14}) show a significant red-to-blue color change within $\sim 200-500$ s after the prompt emission phase.  Moreover, \cite{Morgan14} found a significant decrease of visual extinction $A_{V}$ over a time period of $t\sim 10-100$ s, which is proposed as a first evidence of dust destruction toward GRB 120119A. In particular, late-time observations usually reveal a re-brightening in optical-NIR light curves of GRB afterglows (\citealt{Greiner13}; \citealt{Nardi14}; \citealt{Melandri17}; \citealt{Kann18}). The origin of such an optical re-brightening remains elusive (see e.g., \citealt{Nardi11}). This feature can originate from intrinsic processes related to the central engine of GRBs, external shocks due to interaction of relativistic jet with ambient medium (see e.g., \citealt{Berger03}; \citealt{Melandri17}), or from varying-dust reddening due to dust destruction (\citealt{Draine79}; \citealt{Waxman20}). 

Very recently, \cite{Hoang19} discovered a new dust destruction mechanism called RAdiative Torque Disruption (RATD). The RATD mechanism, which is based on the centrifugal force within rapidly spinning grains spun-up by radiative torques (\citealt{1996ApJ...470..551D}; \citealt{2007MNRAS.378..910L}; \citealt{2009ApJ...695.1457H}), can break a large grain into numerous smaller fragments and require lower radiation intensity than sublimation to be effective. As a result, we expect that the long UV-optical afterglows (up to $10^{5}$ s) after the UV flash can disrupt grains at much later times and farther distances from the central source than prevalent mechanisms. Therefore, the first goal of this paper is to quantify the effect of GRB afterglows on the disruption of dust grains in the surrounding environment and model the time-dependent dust extinction toward GRB afterglows.

Polarimetry is a powerful tool to study the emission mechanism and the geometry of GRB engines. Constraining the geometry of GRB progenitors is particularly important for gravitational wave (GW) astrophysics because GWs are expected to arise from the asymmetric collapse of the iron core of massive stars. Yet, a critical challenge is that the intrinsic polarization of GRB afterglows is uncertain, depending on the geometry and magnetic fields, whereas foreground polarization by circumstellar and interstellar dust in the host galaxy may be dominant. Moreover, numerous observations show time-variation of optical polarization of GRB afterglows (e.g., \citealt{Barth03}), which is explained by means of varying- magnetic fields in the jet (see \citealt{Laskar19} for a review). However, as found in \cite{Giang19} for SNe Ia, we expect that dust polarization due to alignment of dust grains by GRB afterglows would vary with time, which challenges the standard explanation based on the variation of the magnetic fields. Therefore, our second goal is to employ the popular theory of grain alignment and perform detailed modeling of dust polarization arising from grains aligned by GRB afterglows.

The structure of this paper is as follows. We will briefly describe the time-varying luminosity of GRB afterglows and the disruption mechanism in Section \ref{sec:method}. In Sections \ref{sec:GRBext} and \ref{sec:GRBpol}, we present our modeling of time-variation extinction and polarization of GRB afterglows due to grain alignment and disruption by radiative torques. In Section \ref{sec:lightcurve}, we study the effect of grain disruption by RATD mechanism on the observed light curve of GRB afterglows. An extended discussion, including comparison of our theoretical results with observational properties of GRB afterglows, is presented in Sections \ref{sec:discuss}. A summary of our main results is given in Section \ref{sec:sum}.

\section{Radiative Torque Disruption of Grains by GRB afterglows}\label{sec:method}
\subsection{Time-dependent luminosity of GRB afterglows}

The luminosity of GRB afterglows due to the reverse shock (RS) can be described by \cite{Drain02}:
\bea \label{eq:Lnu_RS}
(\nu L_{\rm \nu})_{\rm RS}=L_{0} \frac{(t/t_{0})^{\alpha_{\rm RS}}}{(1+(t/t_{0})^{\alpha_{\rm RS}})^{2}}\left(\frac{h \nu}{13.6 \ev}\right)^{1+\beta},
\ena
where $\alpha_{\rm RS}$ is the RS slope, the spectra index $\beta\sim -0.5$ is usually adopted, $L_{0}$ is the UV-optical luminosity flash at the observed peak brightness $t_{0}$. For GRB 190114C, $\alpha_{\rm RS}=1.5$, $L_{0}$ is normalized to $\sim 2.04\times 10^{50} \erg\s^{-1}$ with a typical observed peak brightness of $t_{0}=10$ s (\citealt{Laskar19}). For $t\gg t_{0}$, $\nu L_{\rm \nu}\propto t^{-1.5}$. It can be seen that even at $t\sim 10^{3}t_{0}=10^{4}~\rm s \sim 3 ~\rm hr$, the luminosity at wavelength $\nu$ still has very high value of $\nu L_{\rm \nu} \sim 10^{11}L_{\odot}$. 

By accounting for the emission by radiative cooling of the forward shock (FS), the luminosity follows a less steep function of time (\citealt{Laskar19}; \citealt{Fraija19}). Therefore, we adopt a function with a shallow slope (\citealt{Fraija19}): 
\bea \label{eq:Lnu_FS}
(\nu L_{\nu})_{\rm FS}=L_{\rm FS} \left(\frac{t}{t_{0}}\right)^{\alpha_{\rm FS}}\left(\frac{h\nu}{13.6\ev}\right)^{1+\beta},
\ena
where $\alpha_{\rm FS}$ is the slope for the FS stage, $L_{\rm FS}$ is the luminosity at a transient phase from RS to FS emission. We adopt $\alpha_{\rm FS}=-0.8$ for GRB 190114C (\citealt{Laskar19}) and get $L_{\rm FS}=9.6\times10^{48}\erg s^{-1}$ at the transient time of $t=1000$ s or $0.01$ day for the case of  $t_{0}=10$ s.

The bolometric luminosity of GRB afterglows can be evaluated as:
\bea
L_{\rm bol}=\int_{1\ev}^{13.6\ev}L_{\rm \nu}d\nu.
\ena

The mean wavelength of the GRB afterglow radiation spectrum can be estimated as:
\bea
\bar{\lambda}=\frac{\int_{1\ev}^{13.6\ev} \lambda L_{\rm \lambda} d\lambda}{\int_{1\ev}^{13.6\ev} L_{\rm \lambda} d\lambda}.\label{eq:lambda_mean}
\ena

Using $\lambda L_{\lambda}=\nu L_{\nu}\propto \nu^{1+\beta}=c^{1+\beta}/\lambda^{1+\beta}$ (see Eqs \ref{eq:Lnu_RS} and \ref{eq:Lnu_FS}), one obtains
\begin{equation*}
\bar{\lambda}=\frac{\int_{1\ev}^{13.6\ev} \lambda^{-(\beta+1)}d\lambda}{\int_{1\ev}^{13.6\ev} \lambda^{-(\beta+2)}d\lambda}=\frac{\beta+1}{\beta}\frac{\lambda^{-\beta}}{\lambda^{-\beta-1}} \Biggr|_{\lambda_{\rm low}}^{\lambda_{\rm up}},
\end{equation*}
where $\lambda_{\rm low}=0.091\mum$ and $\lambda_{\rm up}=1.24\mum$.

Therefore, the mean wavelength becomes
\begin{equation*}
\bar{\lambda}=\frac{\beta+1}{\beta}\frac{\lambda_{\rm up}^{-\beta}-\lambda_{\rm low}^{-\beta}}{\lambda_{\rm up}^{-\beta-1}-\lambda_{\rm low}^{-\beta-1}},
\end{equation*}
which yields $\bar{\lambda}=0.336\mum$ for $\beta=-0.5$.

\subsection{The RATD mechanism}
A dust grain of irregular shape exposed to an anisotropic radiation field experiences radiative torques (\citealt{1976Ap&SS..43..291D}; \citealt{1996ApJ...470..551D}). An analytical of RAdiative Torques (RATs) is developed by \cite{2007MNRAS.378..910L}, and numerical calculations of RATs for many irregular shapes are presented by \cite{Herranen:2019kj}. Experimental test of spin-up by RATs was conducted in \citealt{2004ApJ...614..781A}. \cite{Hoang19} discovered that, in an intense radiation field, the grain rotation rate driven by RATs can be sufficiently large such that induced centrifugal force can disrupt the grain into small fragments, and we termed this mechanism RAdiative Torque Disruption (RATD). A detailed description of the RATD mechanism is presented in \cite{Hoang19}, and its application for type Ia supernovae (SNe Ia) is shown in \cite{Giang19}. Here we only briefly describe the RATD mechanism for the reference.

Let $a$ be the effective grain size defined as the radius of an equivalent spherical grain that has the same volume with an irregular grain. The angular velocity of irregular grains spun-up by RATs is obtained by solving the equation of motion (\citealt{Hoang19}): 
\begin{equation} \label{eq:domega_dt}
\frac{I d\omega}{dt} = \Gamma_{\rm RAT}-\frac{I\omega}{\tau_{\rm damp}},
\end{equation}
where $I=8 \pi \rho a^{5}/15$ is the grain inertia moment with $\rho$ the mass density of grain material, the radiative torque $\Gamma_{\rm RAT}$ is a function of time because of the time-varying luminosity of GRB afterglows, and $\tau_{\rm damp}$ is the characteristic timescale of grain rotational damping induced by gas-grain collisions and IR emission (see \citealt{Hoang19} for details).

A dust grain spinning at angular velocity $\omega$ is disrupted when induced centrifugal stress $S=\rho a^{2}\omega^{2}/4$ exceeds the maximum tensile strength of grain's material, $S_{\rm max}$. The value of $S_{\rm max}$ depends on the grain material, internal structure, and perhaps grain size. It can vary from $S_{\rm max}=10^{11}\erg\cm^{-3}$ for ideal materials, i.e., diamond (\citealt{Draine79}; \citealt{Burke74}) to $S_{\max}\sim 10^{9}-10^{10}\erg\cm^{-3}$ for polycrystalline bulk solid (\citealt{Hoang19}) and $S_{\max}\sim 10^{6}-10^{8}\erg\cm^{-3}$ for composite grains (\citealt{Hoang19b}). In this paper, we take $S_{\rm max}=10^{7}\erg\cm^{-3}$ as a typical value for large grains. Then, the critical angular velocity at which rotational disruption occurs is obtained by setting $S$ equal to $S_{\rm max}$, which yields:
\bea \label{eq:wdisr}
\omega_{\rm disr} &=& \frac{2}{a} \left(\frac{S_{\rm max}}{\rho}\right)^{1/2}\nonumber\\
&\simeq &3.65\times 10^{8}a_{-5}^{-1}\hat{\rho}^{-1/2}S_{\rm max,7}^{1/2}~~~\rm rad\s^{-1},
\ena
where $a_{-5}=a/(10^{-5}\cm)$, $\hat{\rho}=\rho/(3\g\cm^{-3})$, and $S_{\max,7}=S_{\rm max}/(10^{7}\erg\cm^{-3})$.

One can see that for the same density and maximum tensile strength, small grains always need to be spun-up to a higher critical speed than large grains in order to be disrupted by RATD. For example, for $S_{\rm max}=10^{ 7}\erg\cm^{-3}$, grains of $a\sim 0.25\mum$ are disrupted when $\omega \gtrsim 1.46\times 10^{8}$ rad/s, but small grains of $a\sim 0.01\mum$ must be spun-up to $\omega\gtrsim 3.65\times10^{9}$ rad/s. Besides, stronger grains with higher $S_{\rm max}$ are more difficult to disrupt than weak grains with lower $S_{\rm max}$. For instance, the value of $\omega_{\rm disr}$ must be increased to $1.46\times 10^{9}$ rad/s and $3.65\times10^{10}$ rad/s for grains of $0.25\mum$ and $0.01\mum$, respectively, assuming $S_{\rm max}=10^{9}\erg\cm^{-3}$.

Let $U=u_{\rm rad}/u_{\rm ISRF}$ be the strength of a radiation field with $u_{\rm ISRF}=8.64\times10^{-13}\erg\cm^{-3}$ the energy density of the average interstellar radiation field (ISRF) in the solar neighborhood (\citealt{Mathis83}). For strong radiation fields of $U\gg 1$, damping of grain rotation is dominated by IR emission, and the gas damping can be disregarded (see \citealt{Hoang19} for details). Thus, the critical size of rotational disruption, $a_{\rm disr}$, can be given by an analytical formulae (\citealt{Hoang19}; \citealt{Hoang19b}):

\begin{eqnarray} 
\left(\frac{a_{\rm disr}}{0.1~\mu\rm m}\right)^{2.7}&\simeq&2\times 10^{-4}\gamma^{-1}\bar{\lambda}_
{0.5}^{1.7}U_{10}^{-1/3}S_{\rm max,7}^{1/2},\label{eq:adisr}
\end{eqnarray}

where $\gamma$ is the anisotropy degree of the radiation field ($0 \leq \gamma \leq 1)$, $\bar{\lambda}_{0.5} = \bar{\lambda}/(0.5\mum)$, $U_{10}=U/(10^{10}$). The above equation is valid for $a_{\rm disr}\lesssim \bar{\lambda}/1.8$. We also disregard the potential existence of very large grains (size $a\gtrsim 1\mum$) in the surrounding environment, so RATD can disrupt all grains above $a_{\rm disr}$.

One can see that the grain disruption size increases with distance because of the decrease of the radiation energy density as $u_{\rm rad}\propto 1/d^{2}$. For an UV-optical flash of luminosity $L_{\rm bol}\sim 10^{50}\erg\s^{-1}\sim 10^{16}L_{\odot}$, the radiation strength is $U\sim 10^{13} d_{pc}^{-2}$ with $d_{pc}$ the distance given in units of parsec. For weak grains of $S_{\rm max}=10^{7}\erg\cm^{-3}$, Equation (\ref{eq:adisr}) yields $a_{\rm disr}=0.0025\mum$ for $d= 10$ pc and $a_{\rm disr}\sim 0.045\mum$ for $d=100$ pc. For stronger grains of $S_{\rm max}=10^{9}\erg\cm^{-3}$, the disruption size increases to $a_{\rm disr}=0.006\mum$ and $0.01\mum$ at these distances. In realistic situations, the luminosity of GRB afterglows varies with time, as given by Equations (\ref{eq:Lnu_RS}) and (\ref{eq:Lnu_FS}). Thus, the disruption size will be obtained by numerically solving the equation of motion (Eq. \ref{eq:domega_dt}) instead of using Equation (\ref{eq:adisr}).

The disruption time for grains of size $a_{\rm disr}$ can be defined as the time required to spin-up the grains to $\omega_{\rm disr}$:
\begin{eqnarray} \label{eq:tdisr}
t_{\rm disr}&=&\frac{I\omega_{\rm disr}}{dJ/dt}=\frac{I\omega_{\rm disr}}{\Gamma_{\rm RAT}}\simeq 318 \hat{\rho}^{1/2} \bar{\lambda}_
{0.5}^{1.7} \left(\frac{a_{\rm disr}}{0.1~\mu\rm m}\right)^{-0.7}\nonumber\\
&&\times S_{\rm max,7}^{1/2}\left(\gamma U_{10}\right)^{-1} {~\rm s}.
\end{eqnarray}

Equation (\ref{eq:tdisr}) follows that large grains of $a=0.25\mum$ at distance $d$ can be disrupted after disruption time of $t_{\rm disr} = 0.085 d_{pc}^{2} S_{\rm max,7}^{1/2}~\s$. For weak grains of $S_{\rm max}=10^{7}\erg\cm^{-3}$, the disruption time is $t_{\rm disr} \sim 8.5 ~ \rm s$ at $d=10 ~ \rm pc$ and $\sim$ 4 min at 50 pc. For strong grains of $S_{\rm max}\sim 10^{9}\erg\cm^{-3}$, the disruption time increases to $t_{\rm disr}\sim $ 1.5 min and $\sim 36$ min at $d=10$ and 50 pc, respectively.

\section{Extinction of GRB afterglows}\label{sec:GRBext}
In this section, we study the effect of RATD on the extinction of GRB afterglows for an optically thin environment. Thus, all dust grains are exposed to the intrinsic radiation of GRB afterglows.

\subsection{Grain disruption size}
To find the grain disruption size $a_{\rm disr}$ for a variable source like GRB afterglows, we solve Equation (\ref{eq:domega_dt}) to obtain the temporal angular velocity $\omega(t)$ for a range of grain sizes using the luminosity $L_{\rm bol}$ given by Equations (\ref{eq:Lnu_RS}) and (\ref{eq:Lnu_FS}). We then compare $\omega(t)$ with the critical angular velocity of disruption given by Equation (\ref{eq:wdisr}) to obtain $a_{\rm disr}$. The disruption time $t_{\rm disr}$ is also determined.

\begin{figure}[t]
\includegraphics[width=0.5\textwidth]{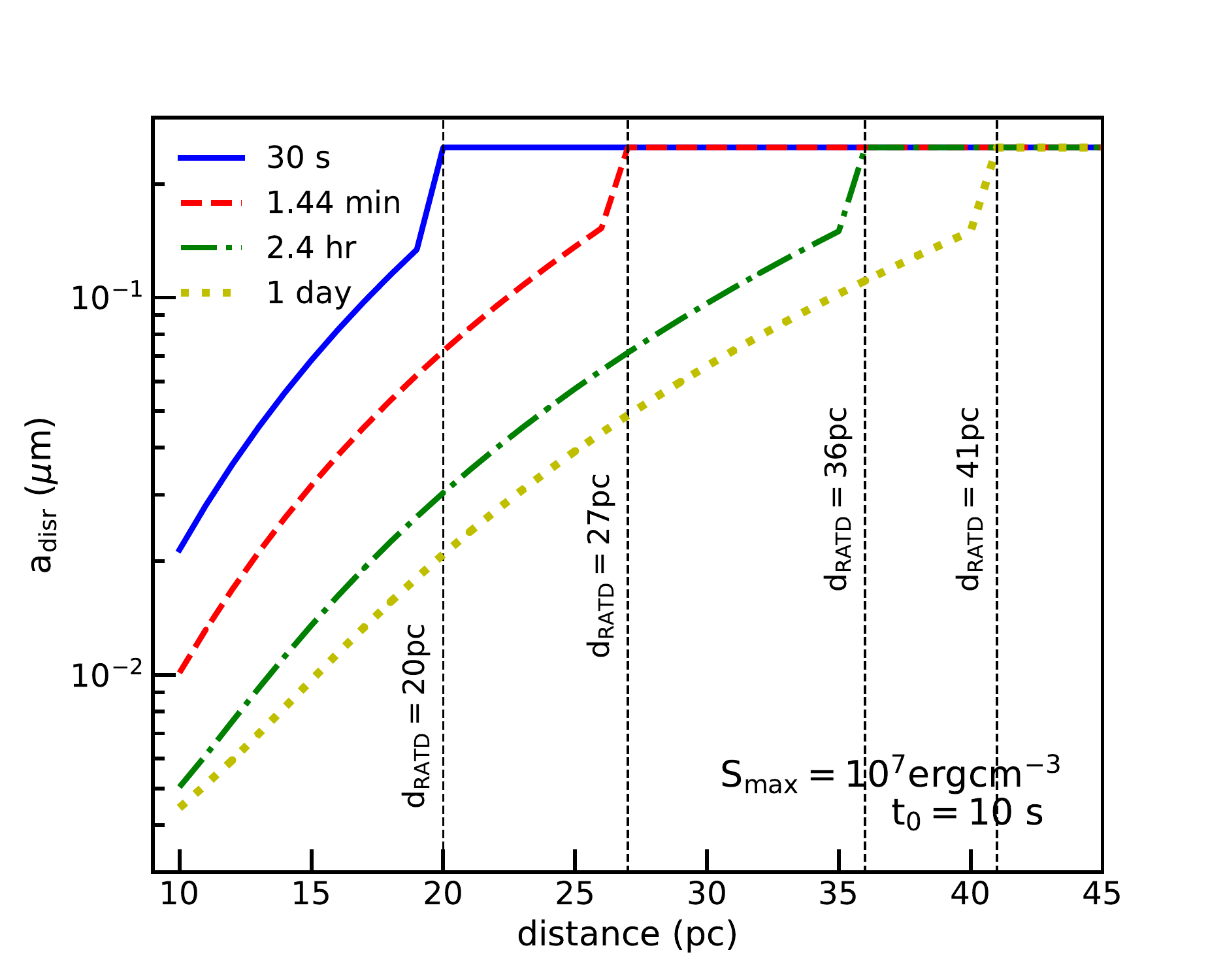}
\includegraphics[width=0.5\textwidth]{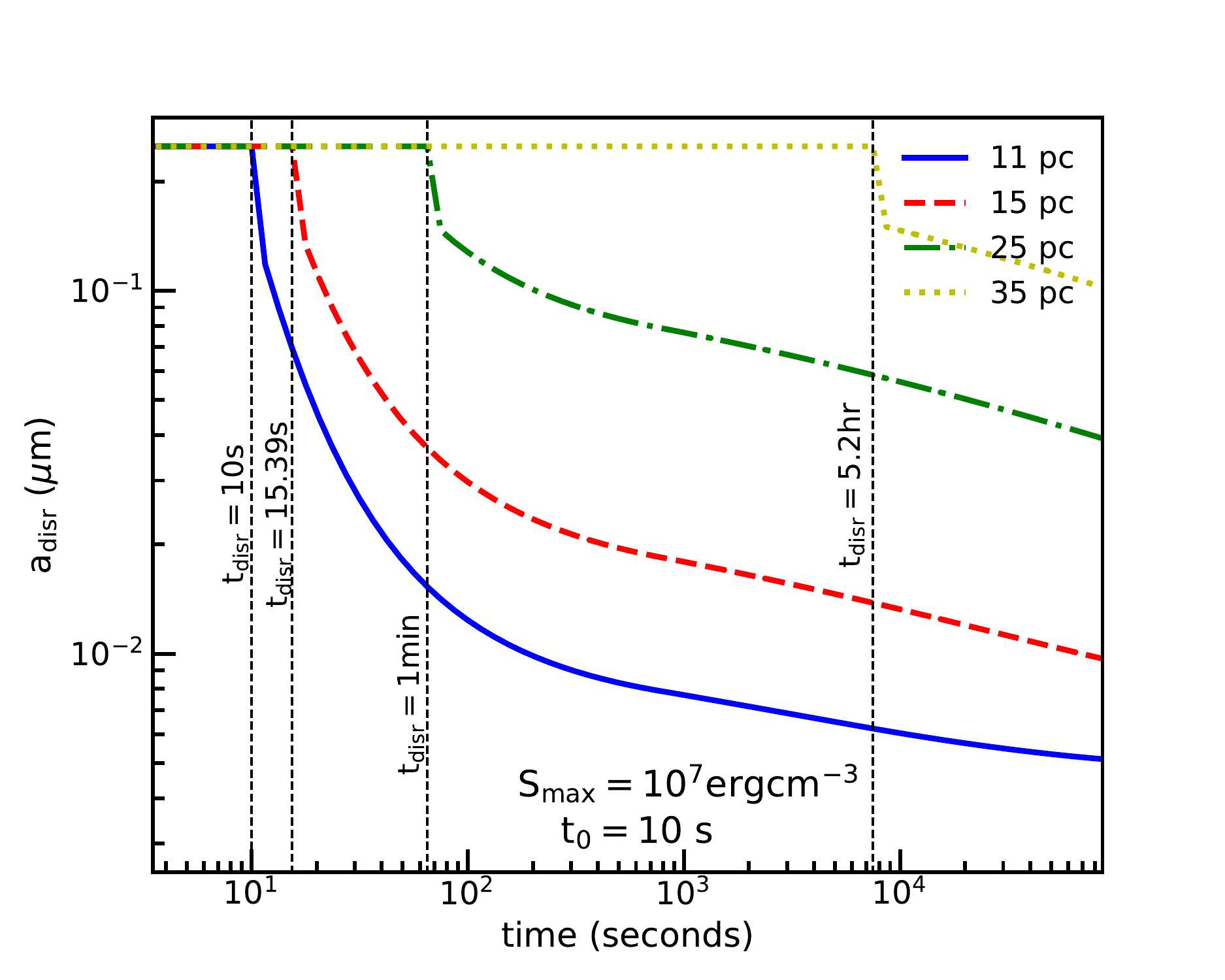}
\caption{Upper panel: grain disruption size by RATD as a function of dust cloud distance at different times since the GRB. Lower panel: variation of grain disruption size with time for different cloud distances from 11 pc to 35 pc. The vertical dotted lines indicate the disruption distance (upper panel) and the disruption time (lower panel). Here the maximum tensile strength $S_{\rm max}=10^{7}\erg\cm^{-3}$ and the peak luminosity of GRB afterglows at $t_{0}=10$ s are assumed.}
\label{fig:adisr_time}
\end{figure}

\begin{figure}[htb!]
\includegraphics[width=0.5\textwidth]{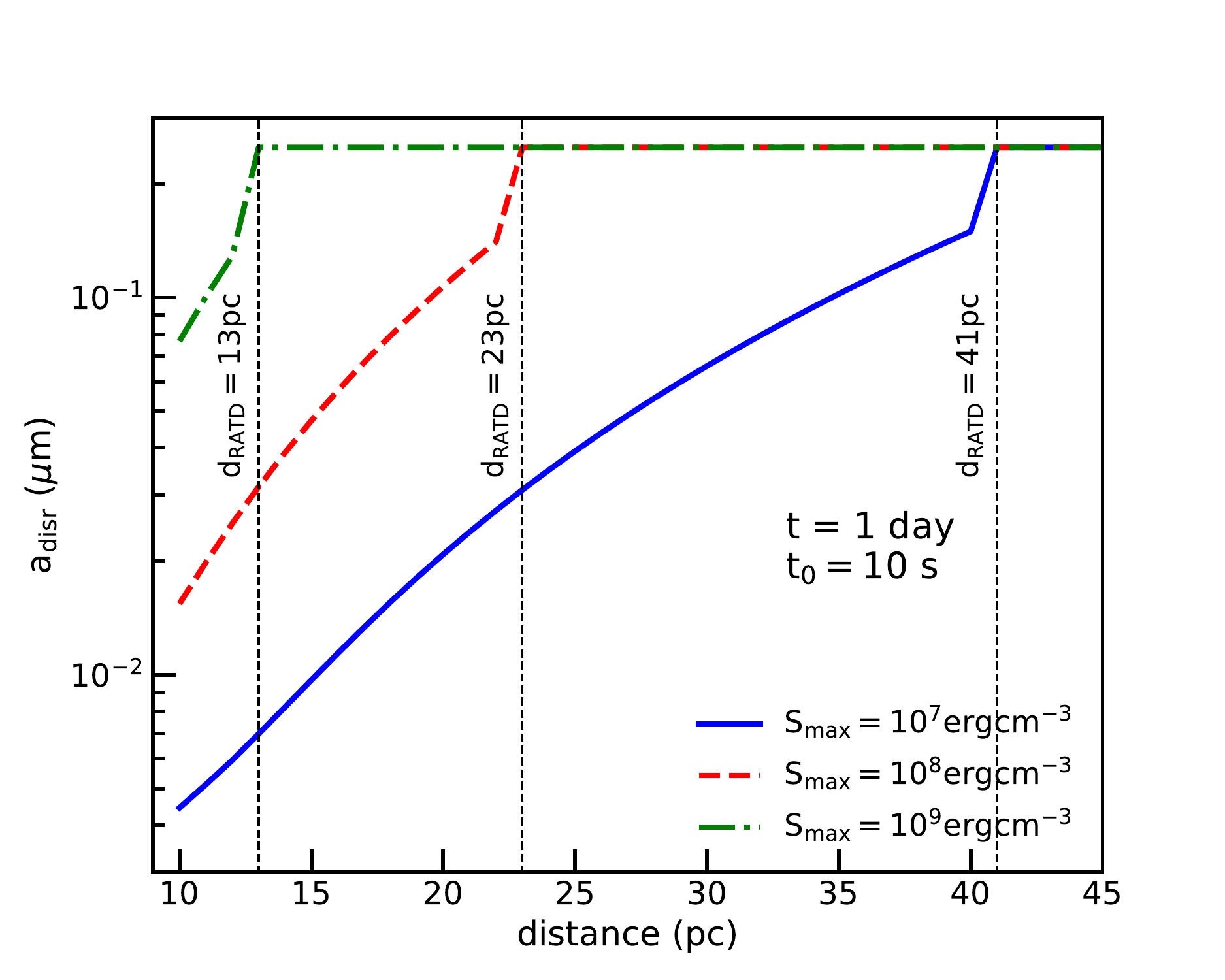}
\includegraphics[width=0.5\textwidth]{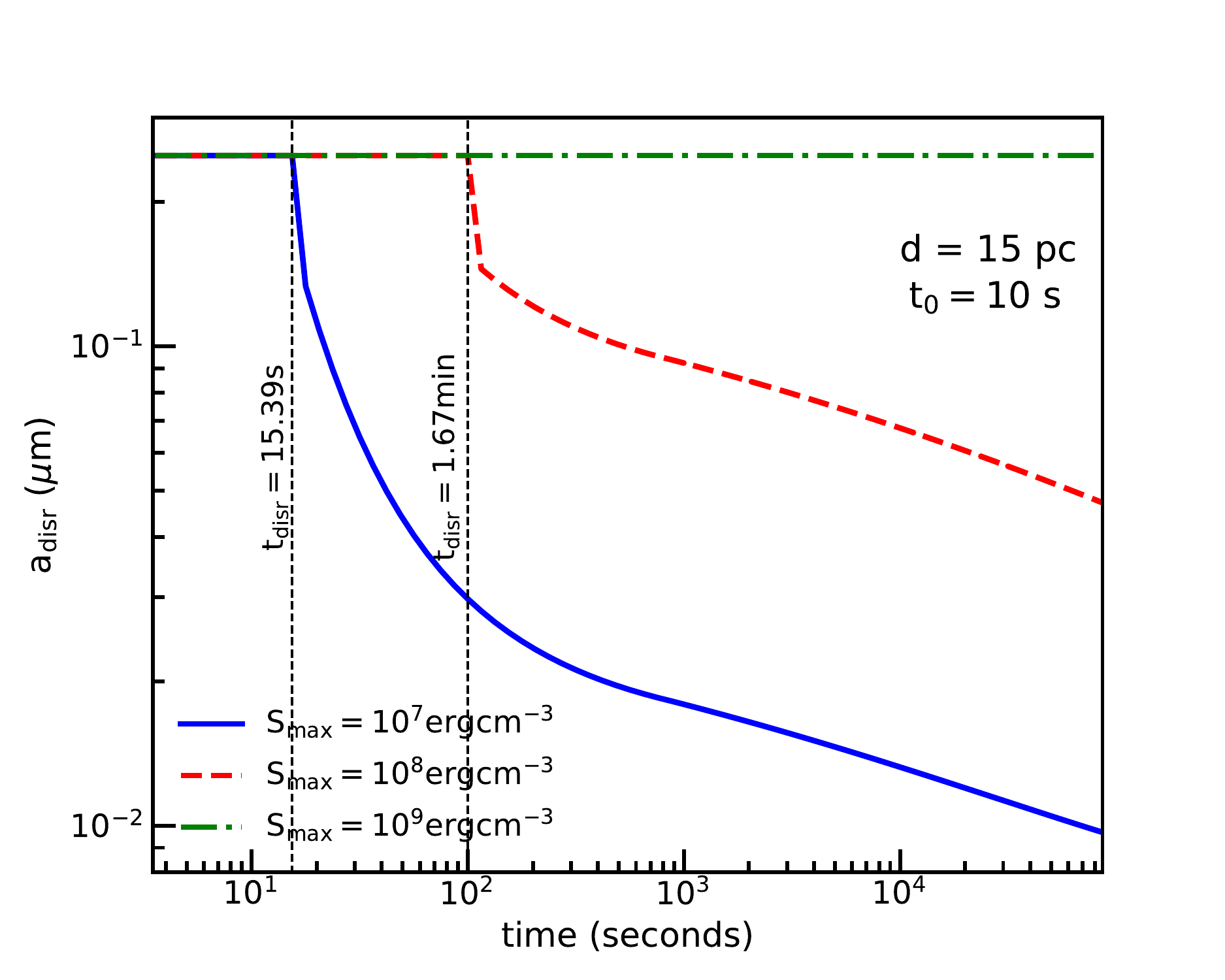}
\caption{Same as Figure \ref{fig:adisr_time} but for the different values of $S_{\rm max}$. The upper panel shows grain disruption size after one day, and the lower panel shows the results for clouds at distance of $15$ pc.}
\label{fig:adisr_Smax}
\end{figure} 

\begin{figure}[htb!]
\includegraphics[width=0.5\textwidth]{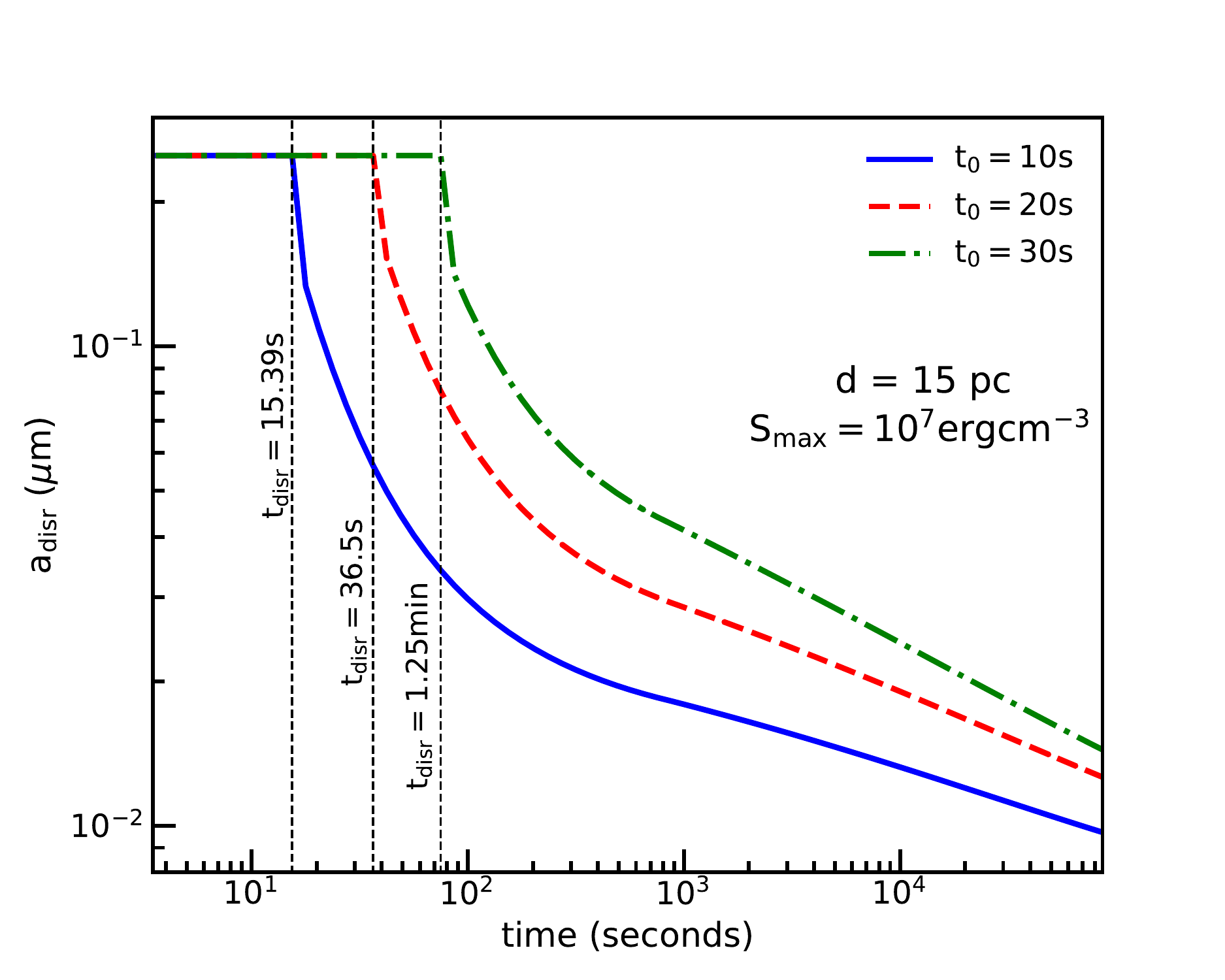}
\caption{Variation of grain disruption size by RATD with time for the different peak luminosity times $t_{0}$, assuming a dust cloud at $d=15$ pc and $S_{\rm max}=10^{7}\erg\cm^{-3}$. RATD occurs earlier (see vertical lines) and $a_{\rm disr}$ can achieve smaller values for smaller $t_{0}$.}
\label{fig:adisr_peaktime}
\end{figure}

Figure \ref{fig:adisr_time} (upper panel) shows the grain disruption size due to the RATD effect as a function of cloud distance at the different time, assuming $S_{\rm max}=10^{7}\erg\cm^{-3}$ and $t_{0}=10$ s. For a given time, the grain disruption size increases with the distance due to the decrease of $u_{\rm rad}$ with distance as $d^{-2}$. Beyond some distance where the energy density becomes insufficiently large to disrupt large grains, we set the grain disruption size $a_{\rm disr}$ to the popular upper limit of the grain size distribution in the interstellar medium of $0.25\mum$ (\citealt{Mathis77}). The distance at which RATD ceases due to the decrease of radiation energy density is called {\it disruption distance}, which defines the active region of RATD. As shown, the disruption distance increases with time, i.e., the curve shifts to larger distances (see vertical dotted lines).
 
Figure \ref{fig:adisr_time} (lower panel) shows the time-variation of grain disruption size for clouds at various distances, using the same value of $S_{\rm max}$ and $t_{0}$ as in the upper panel. 
For a given cloud distance, the disruption size decreases with the irradiation time then cease after a long time. This arises from the fact that large grains that receive stronger RATs can be driven to $\omega_{\rm disr}$ faster than smaller ones. For a dust cloud at $d= 11$ pc, grain disruption begins at $t_{\rm disr}\sim 10$ s, at which $a_{\rm disr}$ starts to decrease from the original value to very small grains of size $a_{\rm disr}\sim 0.005\mum$ after 5 hr. At larger distances of $d=15$ and $25$ pc, grain disruption starts later and the disruption size achieves $a_{\rm disr}\sim 0.015\mum$ and $0.06\mum$ at $t\sim 5$ hours, respectively. At distance $d=35$ pc, grain disruption only occurs after $t\sim 5.2$ hr, and the disruption occurs for large grains of $a>a_{\rm disr}\sim 0.1\mum$ only.

Figure \ref{fig:adisr_Smax} (upper panel) shows the grain disruption size after one day for different cloud distances, assuming $t_{0}=10$ s and different values of $S_{\rm max}$. The active region of RATD reduces from 40 pc for weak grains of $S_{\rm max}=10^{7}\erg\cm^{-3}$ to 25 pc for $S_{\rm max}=10^{8}\erg\cm^{-3}$ and $\sim 10-13$ pc for $S_{\rm max}\geq10^{9}\erg\cm^{-3}$. This arises from the fact that rotational disruption depends closely on the tensile strength of grain materials as shown by Equation (\ref{eq:wdisr}).

Figure \ref{fig:adisr_Smax} (lower panel) shows the grain disruption size versus time for clouds at 15 pc, assuming $S_{\rm max}=10^{7}-10^{9}\erg\cm^{-3}$. Grains with higher $S_{\rm max}$ begin to be disrupted by RATD later compared to weak grains of lower $S_{\rm max}$. For instance, grains with $S_{\rm max}=10^{7}\erg\cm^{-3}$ begin the disruption after $t_{\rm disr}=10$ s and get the grain disruption size of $a_{\rm disr}=0.01\mum$ after one day. However, the disruption time increases to $t_{\rm disr}=15$ s and $a_{\rm disr}=0.05\mum$ for grains with $S_{\rm max}=10^{8}\erg\cm^{-3}$.

Figure \ref{fig:adisr_peaktime} shows the variation of grain disruption size over time for the different values of $t_{0}$, assuming $S_{\rm max}=10^{7}\erg\cm^{-3}$ and cloud distance $d=15$ pc. The grain disruption occurs earlier if the luminosity peaks earlier (smaller $t_{0}$), which arises from the decreases of the luminosity with peak time as $L_{\rm bol}\propto 1/t_{0}$ (see Equation (\ref{eq:Lnu_RS}). For example, with $t_{0}=10$ s, grains of $a=0.25 \mum$ will be disrupted after $t_{\rm disr}=15$ s, and one obtains $a_{\rm disr}=0.02 \mum$ at 1000 s. However, for $t_{0}=30\s$, the $0.25\mum$ grains are disrupted at $t_{\rm disr}\sim$ 1 minutes, and $a_{\rm disr}=0.04\mum$ at 1000 s. 
 
\subsection{Extinction curves} \label{sec:extintion}
To model the extinction of GRB afterglows by intervening dust, we adopt a popular mixed-dust model consisting of astronomical silicate and carbonaceous grains (see \citealt{ Wein01}; \citealt{Drain07}). 

The extinction of GRB afterglows induced by randomly oriented grains in units of magnitude is given by:
\bea
\frac{A(\lambda)}{N_{\rm H}}= \sum_{j=\rm sil,carb} 1.086\int_{a_{\rm min}}^{a_{\rm max}} C_{\rm ext}^{j}(a)\left(\frac{1}{n_{\rm H}}\frac{dn^{j}}{da}\right)da,~~\label{eq:Aext}
\ena
where $a$ is the effective grain size, $dn^{j}/da$ is the grain size distribution of dust component $j$, $C_{\rm ext}$ is the extinction cross-section taken from \cite{Hoang13}, assuming oblate spheroidal grains with axial ratio $r=2$, and $N_{\rm H}$ is the total column density of hydrogen along the line of sight. Here, the maximum grain size $a_{\rm max}={\rm min}(a_{\rm disr},~a_{\rm max,MRN})$ is the upper cutoff of the grain size distribution in the presence of RATD, and $a_{\rm max,MRN}=0.25\mum$ is the upper cutoff of MRN distribution (\citealt{Mathis77}).

Due to the RATD effect, dust extinction given by Equation (\ref{eq:Aext}) is time-dependent because $a_{\rm disr}$ and then $dn/da$ change with time. In order to get insights into the effect of RATD on the time-varying extinction of GRB afterglows, we consider a single slab model, such that the small variation of $a_{\rm disr}$ within the dust cloud can be ignored. Nevertheless, in realistic situations, there may exist several dust clouds between the observer and the GRB afterglow, which is discussed in Section \ref{sec:discuss}.

To model the grain size distribution modified by RATD, we adopt a power law $dn^{j}/da=C_{j} n_{\rm H} a^{\eta}$ with $C_{j}$ the normalization constant of dust component $j$ and $\eta$ the power law slope (\citealt{Mathis77}). For the standard grain size distribution (\citealt{Mathis77}), one has $C_{\rm sil}=10^{-25.11}\cm^{-2.5}$ for silicate grains and $C_{\rm carb}=10^{-25.13}\cm^{-2.5}$ for carbonaceous grains and $\eta=-3.5$. To account for the grain size distribution function modified by RATD, we fix the normalization constant $C$ and change the slope $\eta$. Such a new slope $\alpha$ is determined by the dust mass conservation as given by (see \citealt{Giang19} for more details):
\bea
\int_{a_{\rm min}}^{a_{\rm max}} a^{3} a^{\eta} da =\int_{a_{\rm min}}^{a_{\rm max,MRN}} a^{3} a^{-3.5} da,
\ena
which yields
\begin{align} \label{eq:slope}
\frac{a_{\rm disr}^{4+\eta} - a_{\rm min}^{4+\eta}}{4+\eta}=\frac{a_{\rm max,MRN}^{0.5} - a_{\rm min}^{0.5}}{0.5}.
\end{align}

\begin{figure}[htb!]
\includegraphics[scale=0.45]{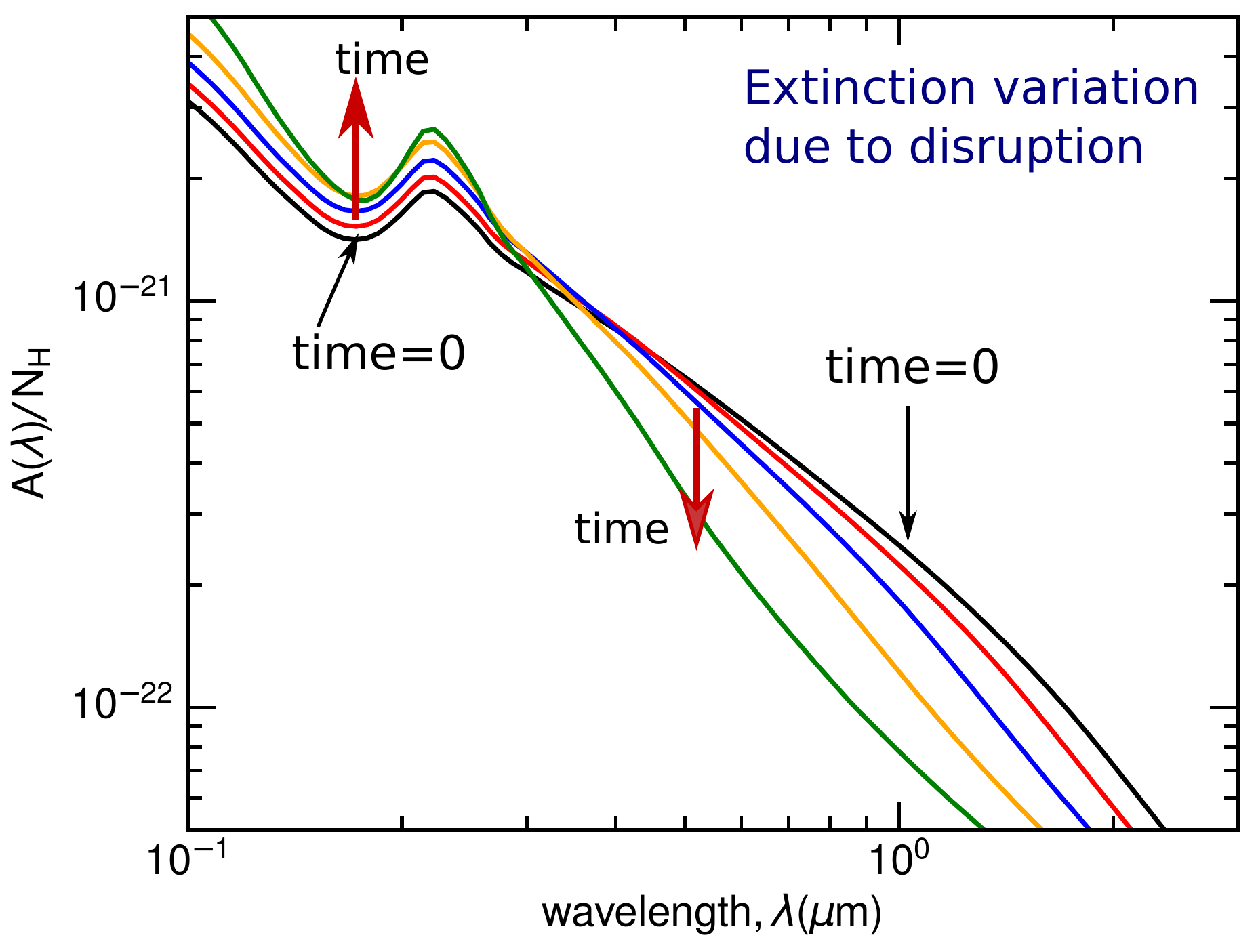}
\caption{A schematic illustration of the time-variation of the extinction curve as a result of RATD. Optical-NIR extinction decreases while UV extinction increases with time due to RATD.}
\label{fig:Aext_time}
\end{figure}

\begin{figure}[htb!]
      \includegraphics[width=0.45\textwidth]{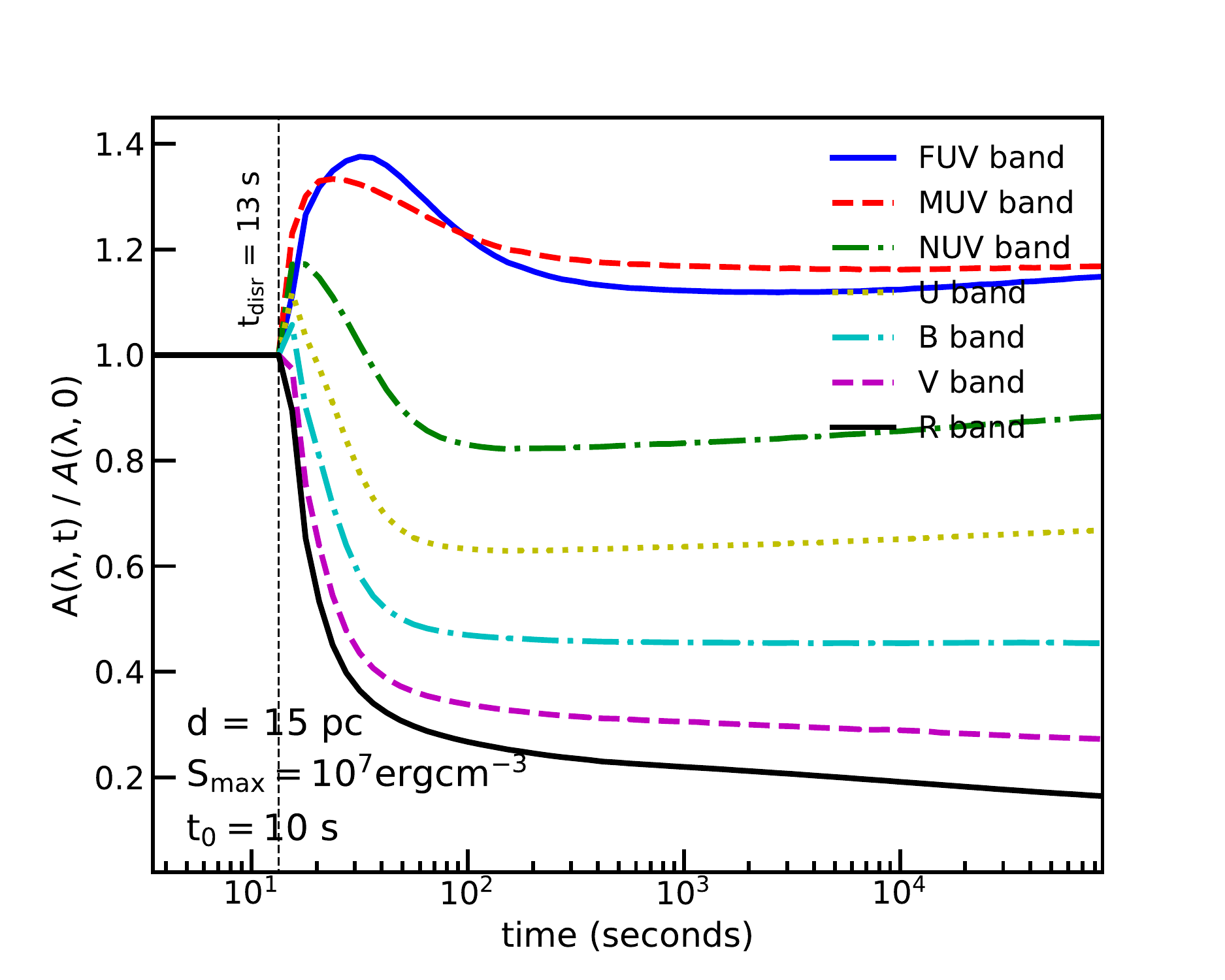}
      \includegraphics[width=0.45\textwidth]{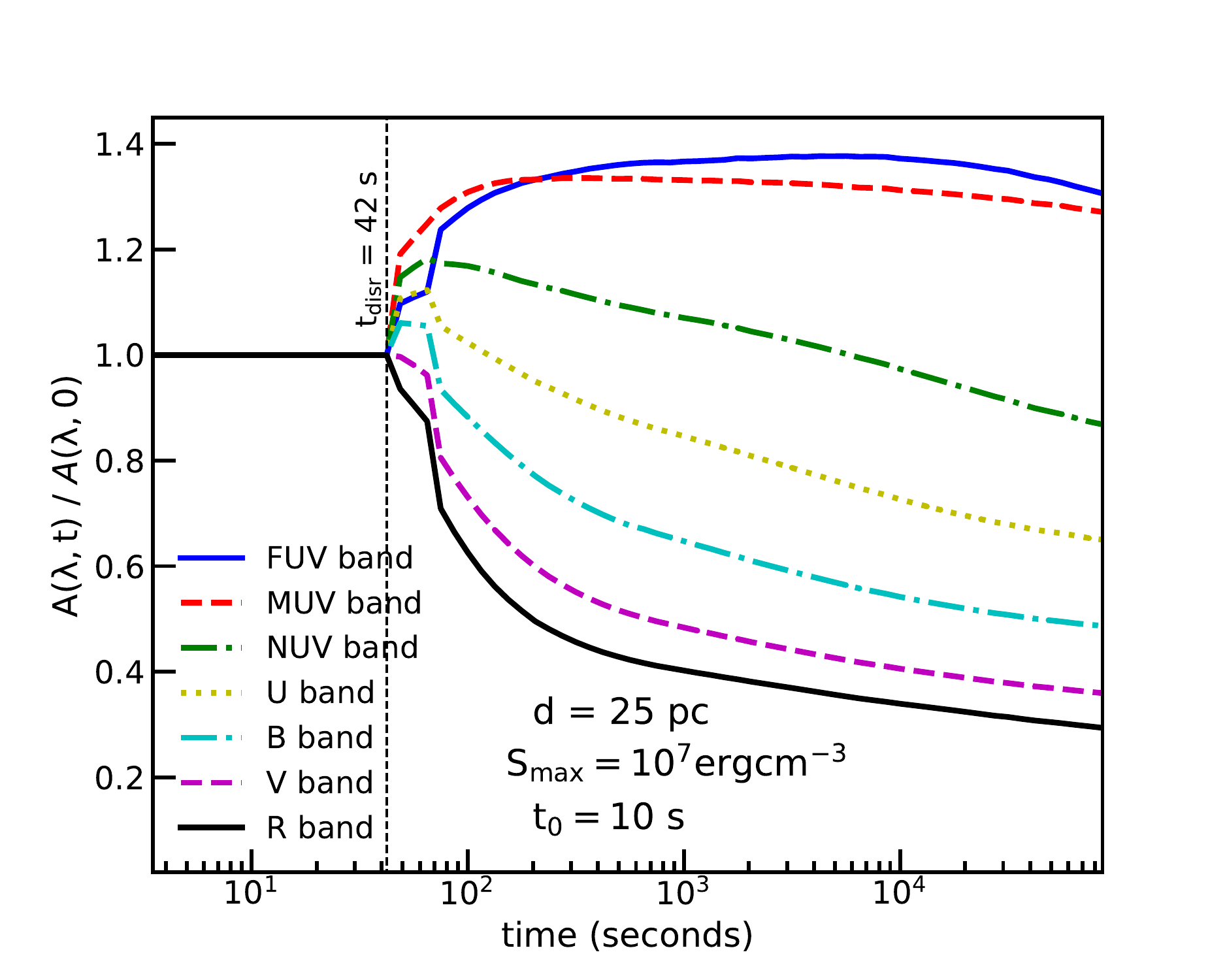}
      \includegraphics[width=0.45\textwidth]{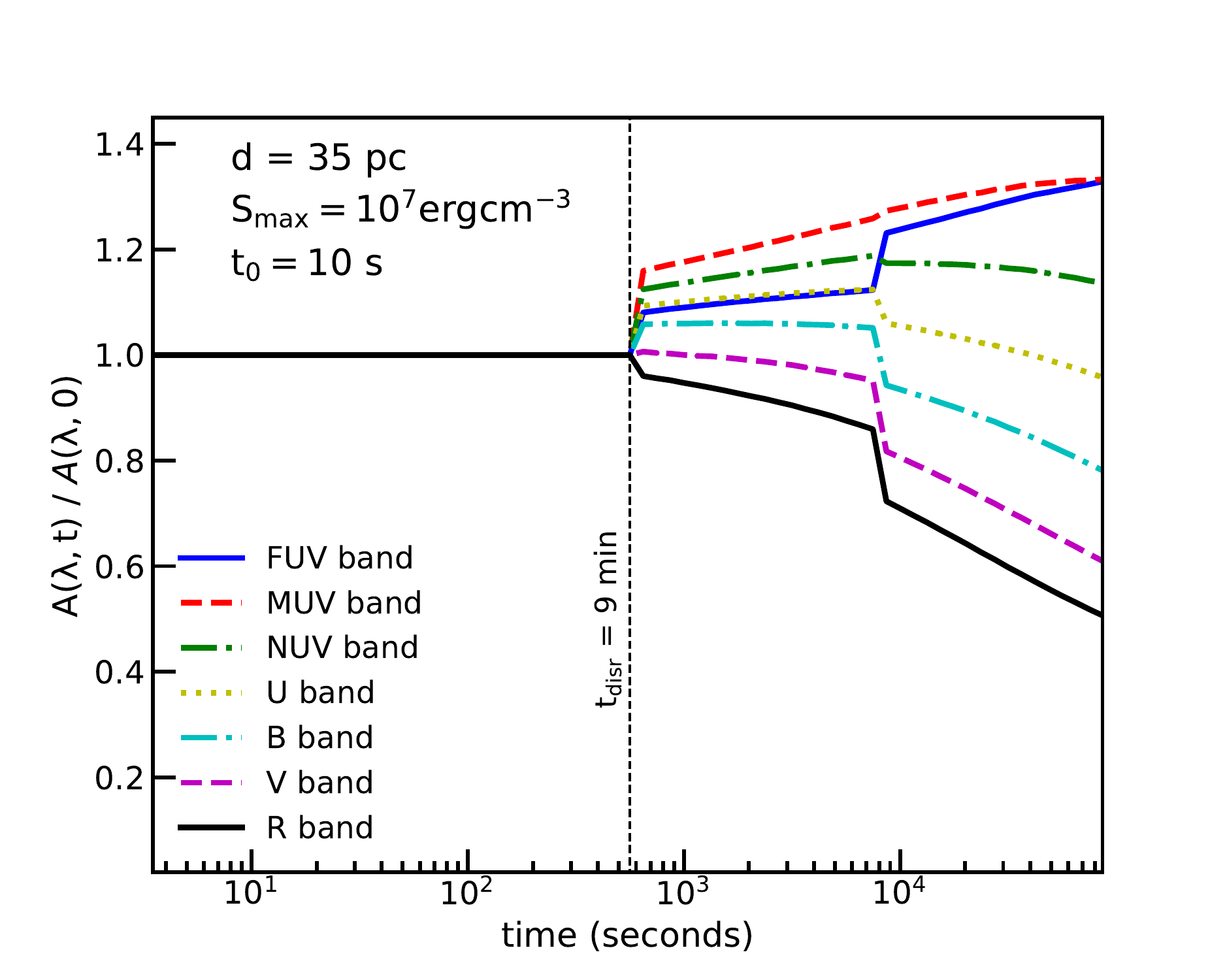}
\caption{Variation of $A(\lambda,t)/A(\lambda,0)$ at different bands with time for dust clouds at 15 pc (upper panel), 25 pc (middle panel) and 35 pc (lower panel) from the radiation source. Dust extinction in all bands start to change when RATD begins at $t=t_{\rm disr}$ (marked by vertical dotted lines). Grain disruption decreases rapidly extinction in UV bands, but increases extinction in MUV and FUV bands.}
\label{fig:Aext_dtime}
\end{figure}
 
Figure \ref{fig:Aext_time} illustrates the time-variation of the extinction curve when the grain size distribution is modified by RATD. The optical to near-infrared (NIR) extinction is seen to decrease gradually with time due to the removal of large grains by RATD. In contrast, ultraviolet (UV) extinction increases due to the enhancement in the abundance of small grains with respect to larger ones. 

Figure \ref{fig:Aext_dtime} shows the variation of $A(\lambda,t)/A(\lambda,0)$ with time from far-ultraviolet (FUV) through optical to (NIR) bands for grains located at distances between $11$ pc to 35 pc from the source,\footnote{Here we start with clouds from 11 pc because thermal sublimation induced by prompt GRB emission can clear out all grains within 10 pc (\citealt{Waxman20}).} assuming $S_{\rm max}=10^{7}\erg\cm^{-3}$ and $t_{0}=10$ s. We choose the central wavelength of the UV range, such as $\lambda=0.15\mum$ for far-UV (FUV) band, $\lambda=0.25\mum$ for mid-UV (MUV) band and $\lambda=0.3\mum$ for near-UV (NUV) band, to study the effect of RATD on the UV extinction. 

Figure \ref{fig:Aext_dtime} shows that dust extinction remains constant for $t\leq t_{\rm disr}$ (before RATD), and changes significantly with time after RATD occurs. One can see that optical-NIR extinction decreases immediately to smaller values because of the quick removal of large grains of size $a\geq 0.1\mum$ by RATD. In contrast, the extinction value in other bands (i.e., U, B and UV bands), first increases due to the enhancement of small grains then decreases later when these small grains are again fragmented into smaller ones. Dust extinction in all bands reaches a saturated value after a long time when RATD ceases. For example, at $d=15$ pc, $A(\lambda)$ stops to change from $\sim 200$ s to one day, which corresponds to the period that $a_{\rm disr}$ only decreases from $0.02\mum$ to $0.01\mum$ (see Figure \ref{fig:adisr_time}, lower panel). For more distant clouds, the variation of dust extinction begins at later times due to larger $t_{\rm disr}$. For instance, the extinction begin to change after $t_{\rm disr}$ = 13 s, 40 s, and 9 minutes for $d=15, 25$ and 35 pc, respectively. 

\subsection{Time-variability of $E(B-V)$ and $R_{\rm V}$}
Using $A(\lambda,t)$ obtained in the previous section, we can calculate the color excess $E(B-V,t)=A_{\rm B}-A_{\rm V}$ and the total-to-selective visual extinction ratio $R_{\rm V}=A_{\rm V}/E(B-V,t)$. Here $A_{\rm V}$ and $A_{\rm B}$ are dust extinction at V and B bands at time $t$.

\begin{figure}[htb!]
\includegraphics[width=0.5\textwidth]{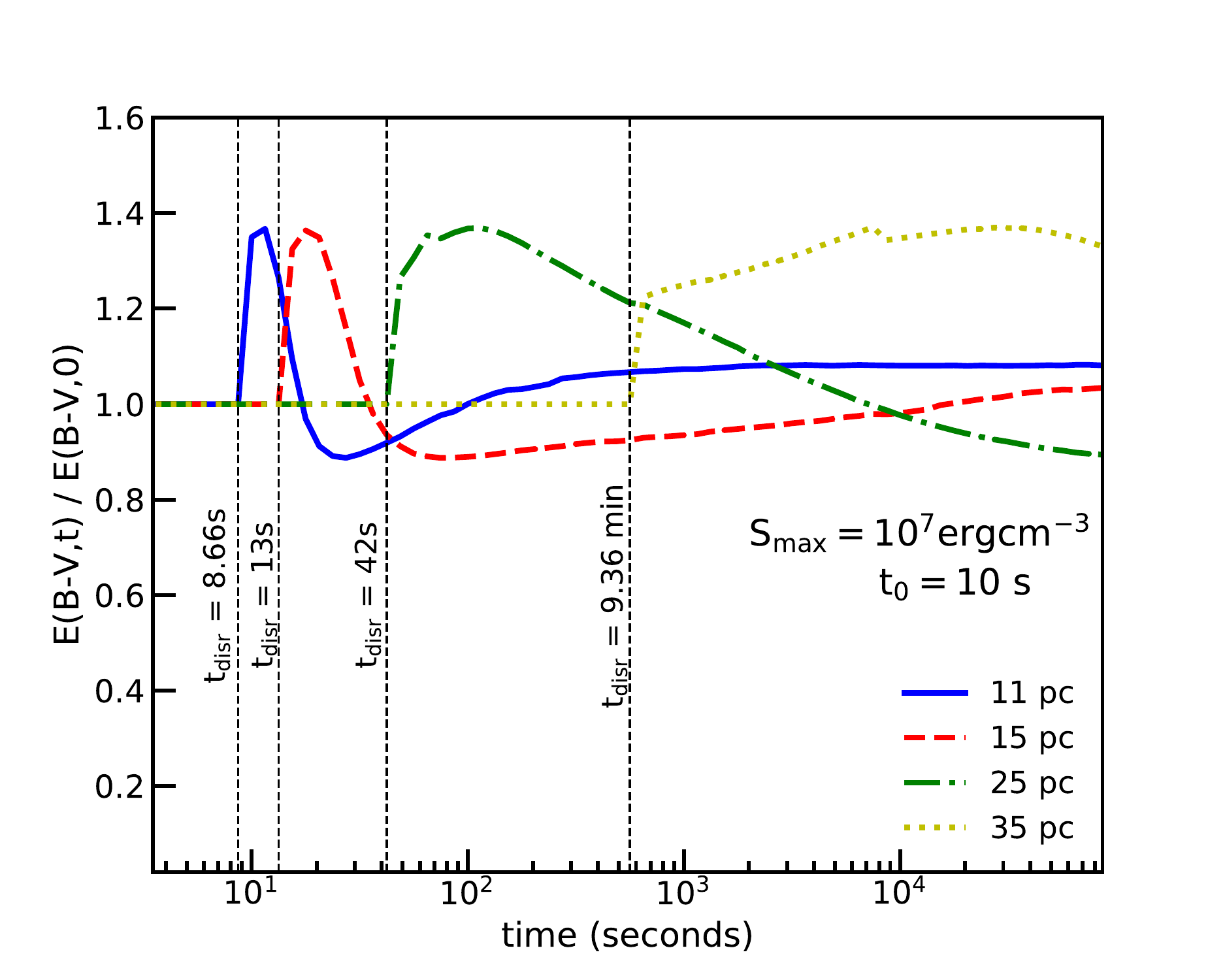}
\includegraphics[width=0.5\textwidth]{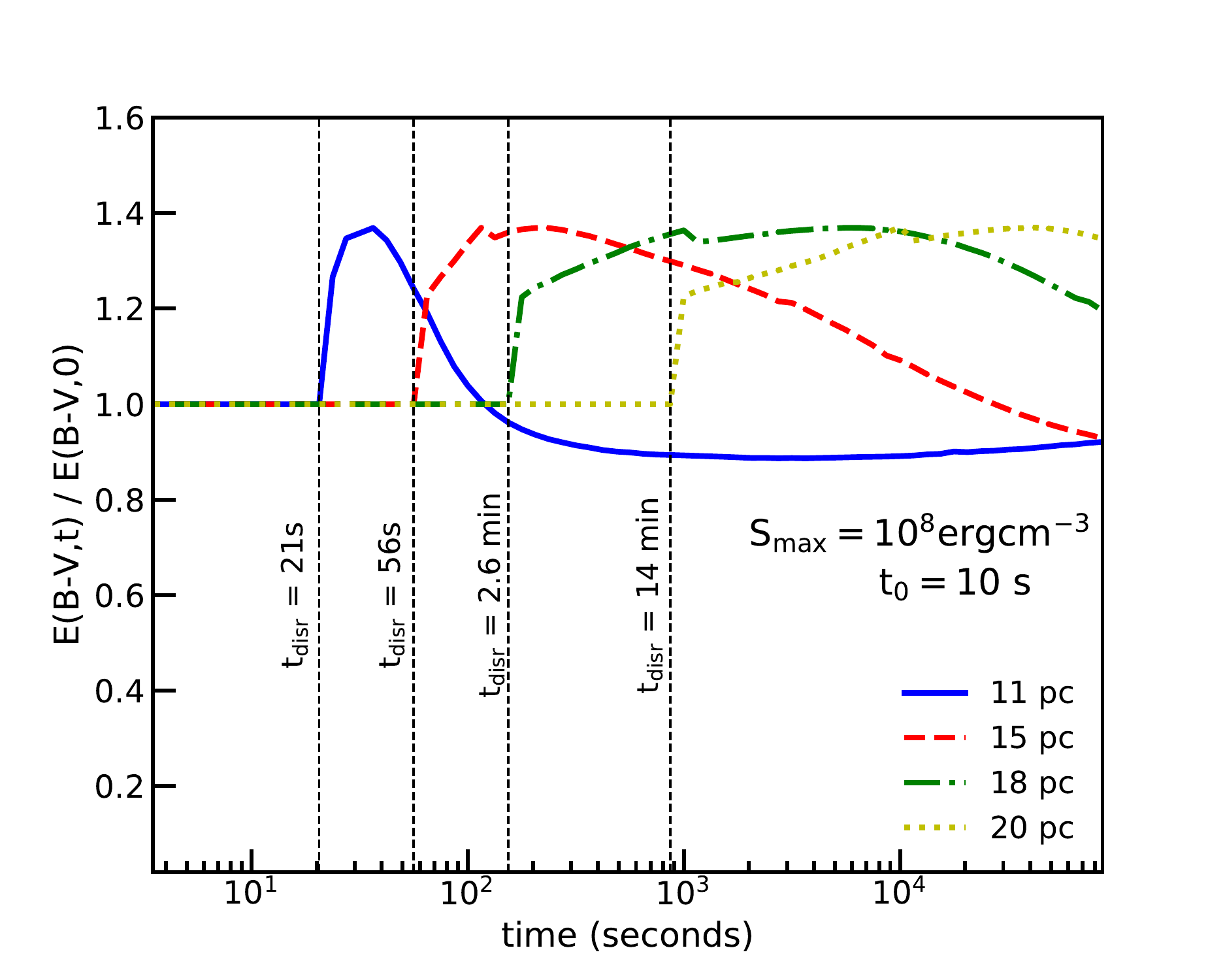}
\caption{Variation of color excess $E(B-V,t)/E(B-V,0)$ with time for different cloud distances, assuming $S_{\rm max}=10^{7}\erg\cm^{-3}$ (upper panel) and $10^{8}\erg\cm^{-3}$ (lower panel). The color excess begins to rise rapidly when grain disruption just starts at $t\sim t_{\rm disr}$ (marked by vertical dotted lines) and then decreases gradually with time.}
\label{fig:EBV}
\end{figure}

Figure \ref{fig:EBV} shows the variation of $E(B-V,t)/E(B-B,t=0)$ with time for different cloud distances from 11 pc to 35 pc, assuming $S_{\rm max}=10^{7}\erg\cm^{-3}$ (upper panel) and $S_{\rm max}=10^{8}\erg\cm^{-3}$ (lower panel) and $t_{0}=10$ s. For a given cloud distance, the color excess remains constant until grain disruption begins at $t\sim t_{\rm disr}$. Subsequently, the ratio increases rapidly and then decreases to a saturated level when RATD ceases. For example, at distance $d=11$ pc, the color excess starts to rise at $t\sim 8.6$ s and declines again to the saturated value at $t\sim 10$ min. The rising stage of $E(B-V)$ is caused by the increase of $A_{\rm B}$ when grain disruption just starts that converts largest grains into smaller ones. Soon after that, these small grains are further disrupted into smaller fragments, both $A_{\rm B}$ and $A_{\rm V}$ decrease (see Figure \ref{fig:Aext_time}), resulting in the decrease of $E(B-V,t)$ with time. Higher tensile strength delays the grain disruption and then the variation of the color excess (see lower panel). For instance, the time-variation of $E(B-V)$ for $d\sim 15 - 25$ pc increases from $13-561$ s for grains with $S_{\rm max}=10^{7}\erg\cm^{-3}$ to $56-840$ s for grains with $S_{\rm max}=10^{8}\erg\cm^{-3}$. 

We note that the amplitude of the $E(B-V,t)$ variation is within $\sim 40\%$, which is different from a large change of $A_{\rm V}$ up to $80\%$ (Figure \ref{fig:Aext_dtime}). This arises from the fact that grain disruption by RATD gradually modifies the grain size distribution.
 
\begin{figure}[htb!]
\includegraphics[width=0.5\textwidth]{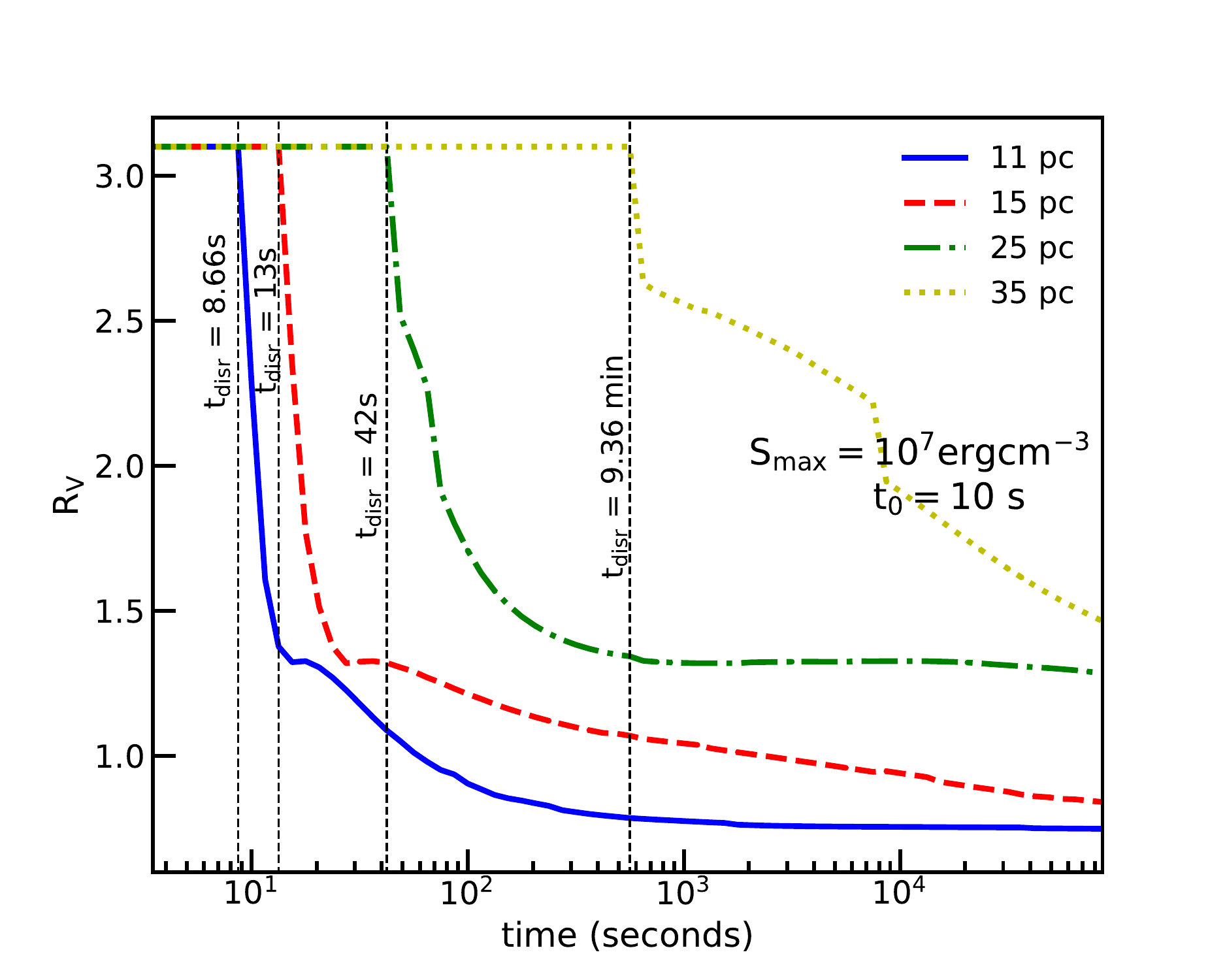}
\includegraphics[width=0.5\textwidth]{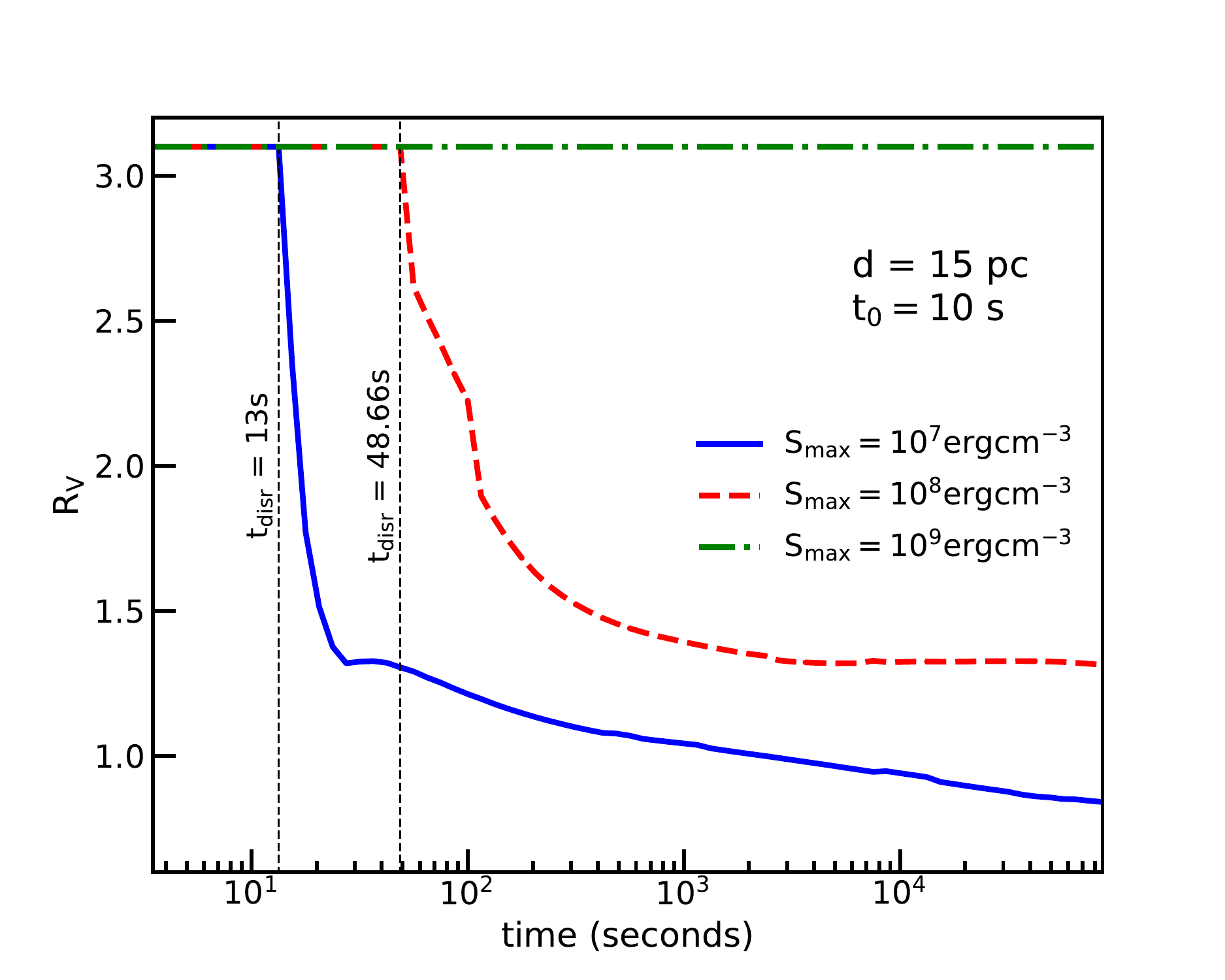}
\caption{Upper panel: variation of total-to-selective visual extinction ratio $R_{\rm V}$ with time for different cloud distances from 11 pc to 35 pc, assuming $S_{\rm max}=10^{7}\erg\cm^{-3}$. Lower panel: variation of $R_{\rm V}$ with time for dust clouds at 15 pc, assuming different values of $S_{\rm max}$. Disruption of grains with $S_{\max}\le 10^{8}\erg\cm^{-3}$ decreases $R_{\rm V}$ with time, but stronger grains of $S_{\rm max} \geq 10^{9}\erg\cm^{-3}$ are not disrupted at this distance and $R_{\rm V}$ remains constant.}
\label{fig:RV_Smax}
\end{figure}
 
Figure \ref{fig:RV_Smax} (upper panel) shows the variation of $R_{\rm V}$ with time for different cloud distances from 11 pc to 35 pc, assuming $S_{\rm max}=10^{7}\erg\cm^{-3}$ and $t_{0}=10$ s. For a given distance, one can see that $R_{\rm V}$ begins to decrease rapidly from its original value of $3.1$ given by standard dust in ISM at $t=t_{\rm disr}<10$ min to smaller values of $R_{\rm V}\sim 0.5 -1.5$ due to RATD. The final values $R_{\rm V}$ is larger for grains located further away from the source.
 
Figure \ref{fig:RV_Smax} (lower panel) shows the time-variation of $R_{\rm V}$ during one day for clouds at 15 pc and different tensile strengths. The value of $R_{\rm V}$ decreases quickly with time for weak grains of $S_{\rm max}=10^{7}\erg\cm^{-3}$ and $S_{\rm max}=10^{8}\erg\cm^{-3}$, but $R_{\rm V}$ does not change for strong grains of $S_{\rm max}\geq 10^{9}\erg\cm^{-3}$.
 
\section{Polarization of GRB afterglows}\label{sec:GRBpol}

\subsection{Grain alignment size} \label{sec:align}

Following the RAdiative Torque (RAT) mechanism (see \citealt{Ander15} and \citealt{Laza15} for recent reviews), dust grains subject to the GRB afterglow can be aligned with the ambient magnetic field when they can keep its orientation in the radiation field by being spun-up to the suprathermal speed.\footnote{Grains may be aligned with the long axis perpendicular to the radiation direction in the intense radiation field (\citealt{2007MNRAS.378..910L}). However, here we stick to the traditional mechanism of grain alignment with the magnetic field.} The suprathermal rotation condition is approximately given by (\citealt{Hoang:2008}; \citealt{Hoang16}):
\bea \label{eq:walign}
\omega_{\rm RAT} \geq 3 \omega_{T},
\ena
where $\omega_{T}$ is the thermal angular velocity of dust grains at gas temperature $T_{\rm gas}$:
\bea \label{eq:wther}
\omega_{T} &=& \sqrt\frac{2 k T_{\rm gas}}{I}\\ \nonumber
&& \simeq 2.3\times 10^{5} \hat{\rho}^{-1/2} a_{-5}^{-5/2} \left(\frac{T_{\rm gas}}{100\K}\right)^{1/2}~\rm rad\s^{-1},
\ena
where $k$ is the Boltzmann constant. 

For a given cloud with gas temperature $T_{\rm gas}$, small grains have a higher suprathermal threshold than large grains. As a result, they require higher radiation energy (i.e., closer clouds) to be efficiently aligned by RATs. 

Based on Equation (\ref{eq:walign}), the grain size at $\omega_{\rm RAT} = 3 \omega_{T}$ is defined as the critical size of grain alignment, $a_{\rm align}$. All grains larger than $a_{\rm align}$ are assumed to be perfectly aligned (\citealt{Hoang16}). Following \cite{Hoang17}, the grain alignment size is given by: 
\bea \label{eq:align}
\left(\frac{a_{\rm align}}{0.1 \mum}\right)^{4.2}& \simeq &1.4\times 10^{-5} \hat{\rho}^{-1/2} \gamma^{-1} \bar{\lambda}_{0.5}^{1.7} U_{6}^{-1/3}\nonumber\\
&&\times \left(\frac{T_{\rm gas}}{100\K}\right)^{1/2},
\ena
where the dominance of IR damping over gas damping is used, which is valid for the intense radiation field of GRB afterglows. Above, we disregard the dependence of the rotation rate spun-up by RATs on the angle between the radiation direction and the magnetic field (\citealt{Hoang:2009}). Accounting for that effect would reduce the value of $\omega_{\rm RAT}$ and $a_{\rm align}$ would become larger. However, the time-variability of $a_{\rm align}$ would not change significantly because it is determined by the varying luminosity of GRBs.

With the same assumption of GRB afterglows in Section \ref{sec:method}, one can find that the grain alignment size increases with increasing cloud distance and gas temperature. For instance, the grain alignment size will be $a_{\rm align}=0.0023 \mum$, $0.003\mum$ and $0.0034\mum$ for cloud at 10 pc, 50 pc, and 100 pc, respectively, assuming $T_{\rm gas} = 100~\rm K$. It will increase to $a_{\rm align}=0.0029\mum$, $0.0037\mum$ and $0.0042\mum$, respectively for $T_{\rm gas}=500$ K. 

Initially ($t=0$ s), grains are aligned by the average interstellar radiation field ($\gamma=0.1$ and $U=1$) with $a_{\rm align}\sim 0.051\mum$ (see e.g., Eq.\ref{eq:align}). The alignment time $t_{\rm align}$ is defined as the time required for grains to be spun-up to suprathermal rotation intense radiation field of GRB afterglows (\citealt{Hoang17}: \citealt{Giang19}):

\bea \label{eq:talign}
t_{\rm align} &=& \frac{3 I \omega_{T}}{dJ/dt} \simeq 0.6\hat{\rho}^{1/2} \left(\frac{T_{\rm gas}}{100\K}\right)^{1/2} \left(\frac{a_{\rm align}}{0.1 \mum}\right)^{-2.2}\nonumber\\
&&\times \left(\frac{\bar{\lambda}_{0.5}^{1.7}}{\gamma U_{10}}\right) \s.
\ena

We obtain the alignment time of $t_{\rm align}\simeq 0.0012 d_{\rm pc}^{2} (T_{\rm gas}/100\K)^{1/2}$ s, which is very short for grains at $d\sim 1$ pc from the source. For different clouds at 10 pc, 50 pc and 100 pc, we get $t_{\rm align}=0.1213$ s, 3.4 s and 13 s, respectively if clouds has $T_{\rm gas}=100$ K. For higher gas temperature $T_{\rm gas}=500\K$, $t_{\rm align}$ increases to 0.3 s, 7.5 s and 30 s. However, we note that these calculations above assume the constant luminosity of GRB afterglows, which overestimates the value of $a_{\rm align}$ and $t_{\rm align}$ than the real case due to the time-varying luminosity of GRBs. From Equations (\ref{eq:talign}) and (\ref{eq:tdisr}), it follows that the alignment time is much smaller than the disruption time. This is obvious because RAT alignment can occur at rotation rates much lower than RATD.
 
\begin{figure}[htb!]
\includegraphics[width=0.5\textwidth]{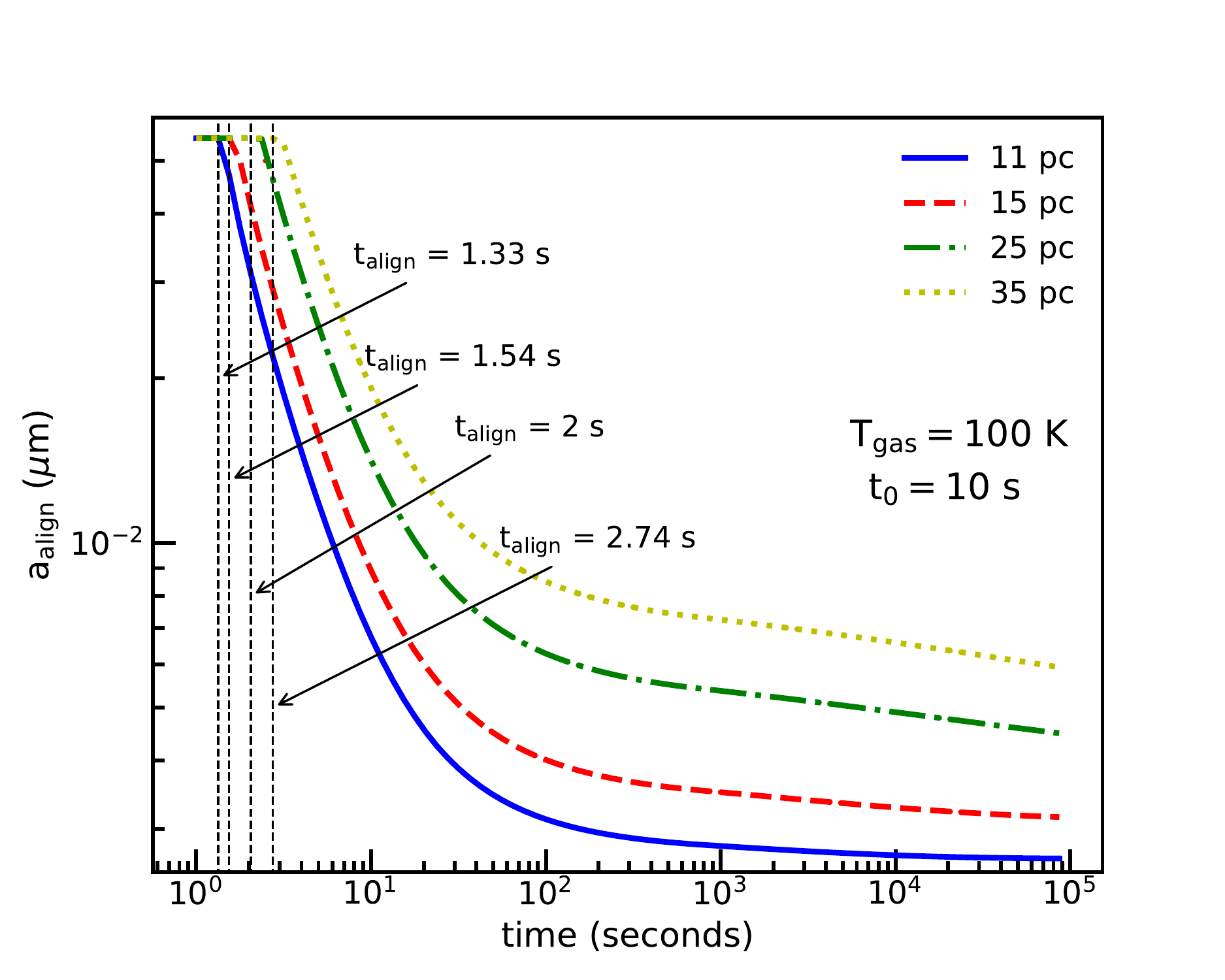}
\caption{Variation of grain alignment size induced by RATs from GRB afterglows with time for different cloud distances from 11 pc to 35 pc, assuming gas temperature $T_{\rm gas}=100$ K. Alignment sizes rapidly decreases with time due to increasing rotation rate from RATs.}
\label{fig:align_time}
\end{figure}

Figure \ref{fig:align_time} shows the variation of the grain alignment size $a_{\rm align}$ during the first day for different cloud distances from 11 pc to 35 pc, assuming $T_{\rm gas}=100~\rm K$. The alignment size first remains constant at $a_{\rm align}\approx 0.055\mum$ given by the alignment of the average interstellar radiation until $t_{\rm align}$. After that, it decreases rapidly to smaller values until $t\sim 100$ s and then slows down later due to the decrease of the radiation intensity. For more distant clouds, $t_{\rm align}$ becomes larger and $a_{\rm align}$ starts to decrease later due to a lower radiation intensity. For example, the alignment time increases from $\sim$ 1.33 s for grains at 11 pc to $\sim$ 3 s for grains at 35 pc. Also, after one day, RATs can align small grains of $a\sim 0.003\mum$ and $0.006\mum$ for the two above distances, respectively. 

\subsection{Polarization curves} \label{sec:pola}
 
Observations (\citealt{Chiar06}) and theoretical studies (\citealt{Hoang16}) reveal that carbonaceous grains are unlikely aligned with the ambient magnetic field due to their diamagnetic properties (see \citealt{Laza15} for a review). Therefore, we assume that carbonaceous grains are randomly oriented, and only silicate grains can be aligned by RATs. The degree of polarization (in the unit of $\%$) of an GRB afterglow induced by differential extinction by aligned grains along the line of sight is computed by
\bea \label{eq:Plamda}
\frac{P(\lambda)}{N_{\rm H}}= 100 \int_{a_{\rm min}}^{a_{\rm max}} \frac{1}{2}C_{\rm pol}^{\rm sil}(a)f(a)\cos^{2}\zeta
\left(\frac{1}{n_{\rm H}}\frac{dn^{\rm sil}}{da}\right)da,~~~\label{eq:Plam}
\ena
where $C_{\rm pol}^{\rm sil}$ is the polarization cross-section, $f(a)$ is the effective degree of grain alignment for silicate grains of size $a$ (hereafter alignment function), and $\zeta$ is the angle between the magnetic field and the plane of the sky (see \citealt{Hoang17}). We take $C_{\rm pol}$ computed for different grain sizes and wavelengths from \cite{Hoang13}. 

We model the size-dependence degree of grain alignment by RATs as follows:
\bea
f(a)=1-\exp \left[-\left(\frac{0.5a}{a_{\rm align}}\right)^{3}\right],\label{eq:fali}
\ena
where $a_{\rm align}$ is given by Equation (\ref{eq:align}) (\citealt{Hoang14}; \citealt{Hoang15a}). This alignment function returns $f(a)=1$ (i.e., the perfect alignment) for large grains of size $a\gg a_{\rm align}$ and approximates the size-dependence alignment degree computed from simulations for grains with enhanced magnetic susceptibility by \cite{Hoang16}. Here we take $a_{\rm max} = \min(a_{\rm disr}, a_{\rm max, MRN})$ and $dn/da$ as used for dust extinction.

Above, we have assumed that small grains (i.e., $a<0.05\mum$) can be perfectly aligned with the magnetic field if they can be spun-up to suprathermal rotation by RATs. However, such small grains may not have iron inclusions (\citealt{1986ApJ...308..281M}), and the degree of grain alignment induced by only RATs for ordinary paramagnetic grains may not be perfect if RAT alignment lacks high-J attractor points (\citealt{Hoang16}). Due to uncertainty in magnetic properties of dust grains, our theoretical predictions in this section are considered upper limit of dust polarization. 

\begin{figure}[htb!]
      \includegraphics[width=0.5\textwidth]{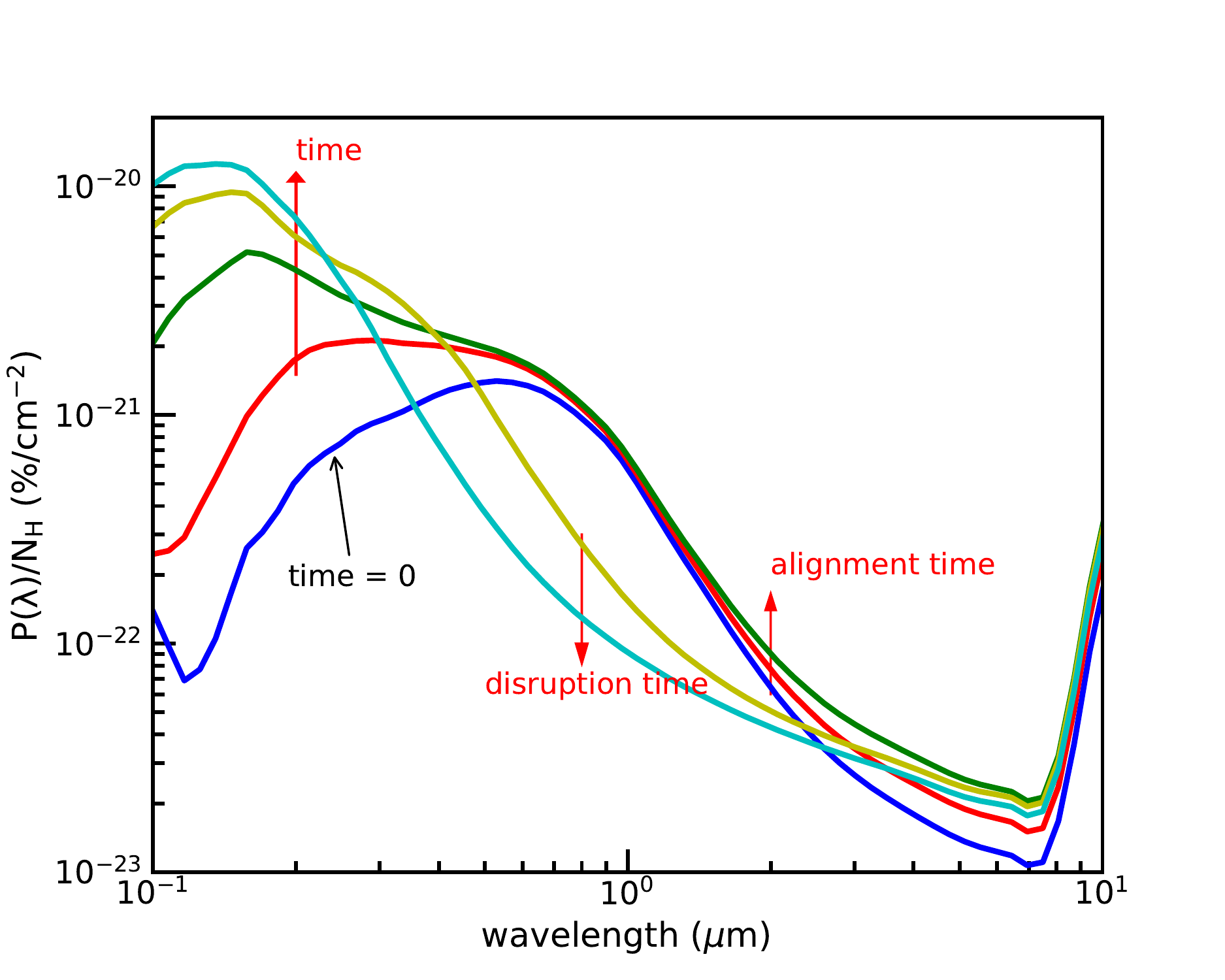}
\caption{A schematic illustration for the variation of the polarization curve due to RAT alignment and RATD. The polarization degree at short wavelengths (UV) increases with time due to enhanced alignment of small grains, while optical-NIR polarization decreases due to the removal of large grains by RATD.}
\label{fig:Polabs_time}
\end{figure}

Figure \ref{fig:Polabs_time} illustrates the general variation of the polarization curve with time as a result of RAT alignment and RATD. As soon as the alignment by RATs starts to occur, the degree of polarization increases and the peak wavelength shifts to shorter wavelengths due to an enhanced alignment of small grains. When RATD begins, the optical-NIR polarization is significantly reduced but UV polarization increases due to the conversion of large grains into small grains. As a result, the polarization curve will narrow with time.

\begin{figure}
\includegraphics[width=0.45\textwidth]{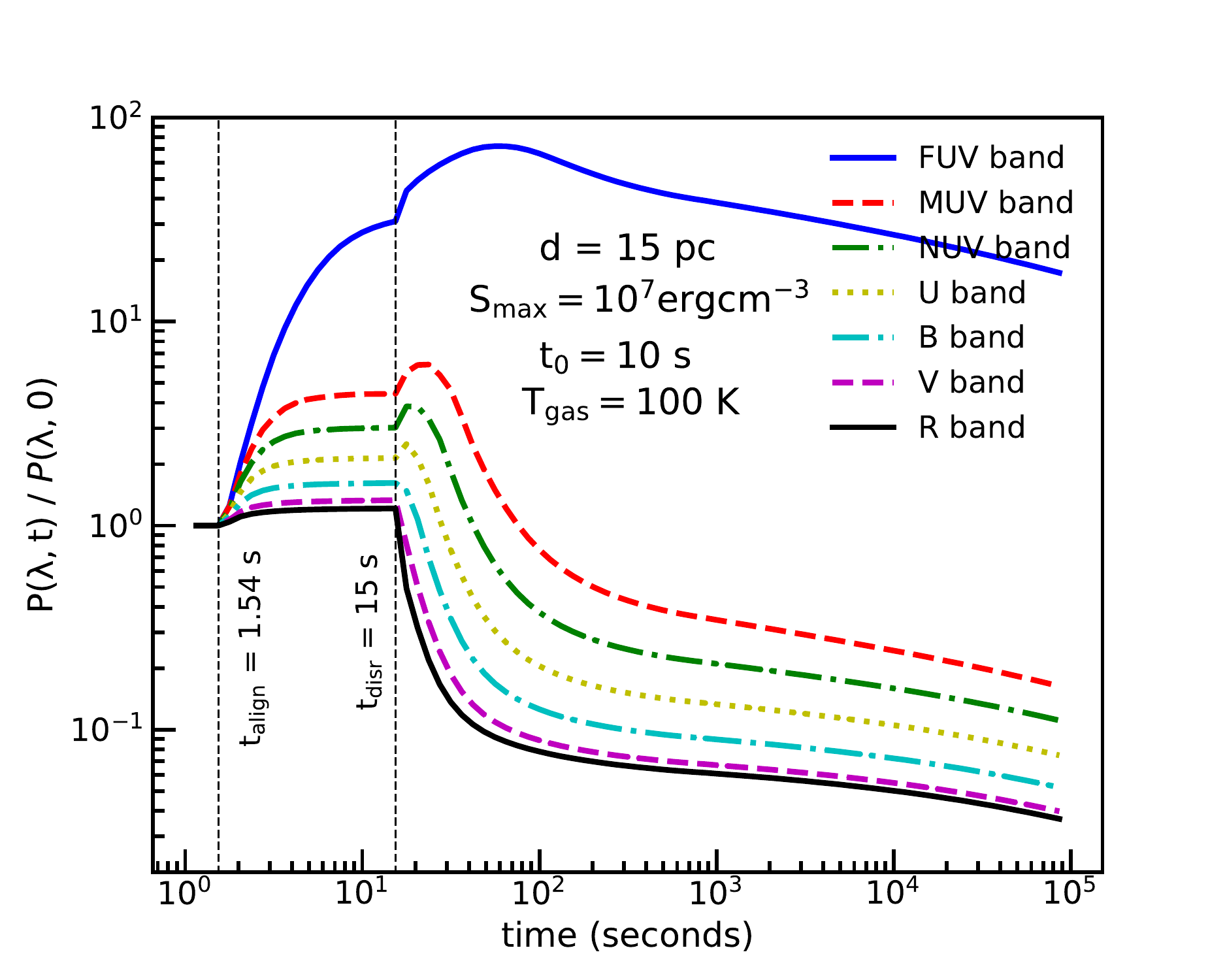}
\includegraphics[width=0.45\textwidth]{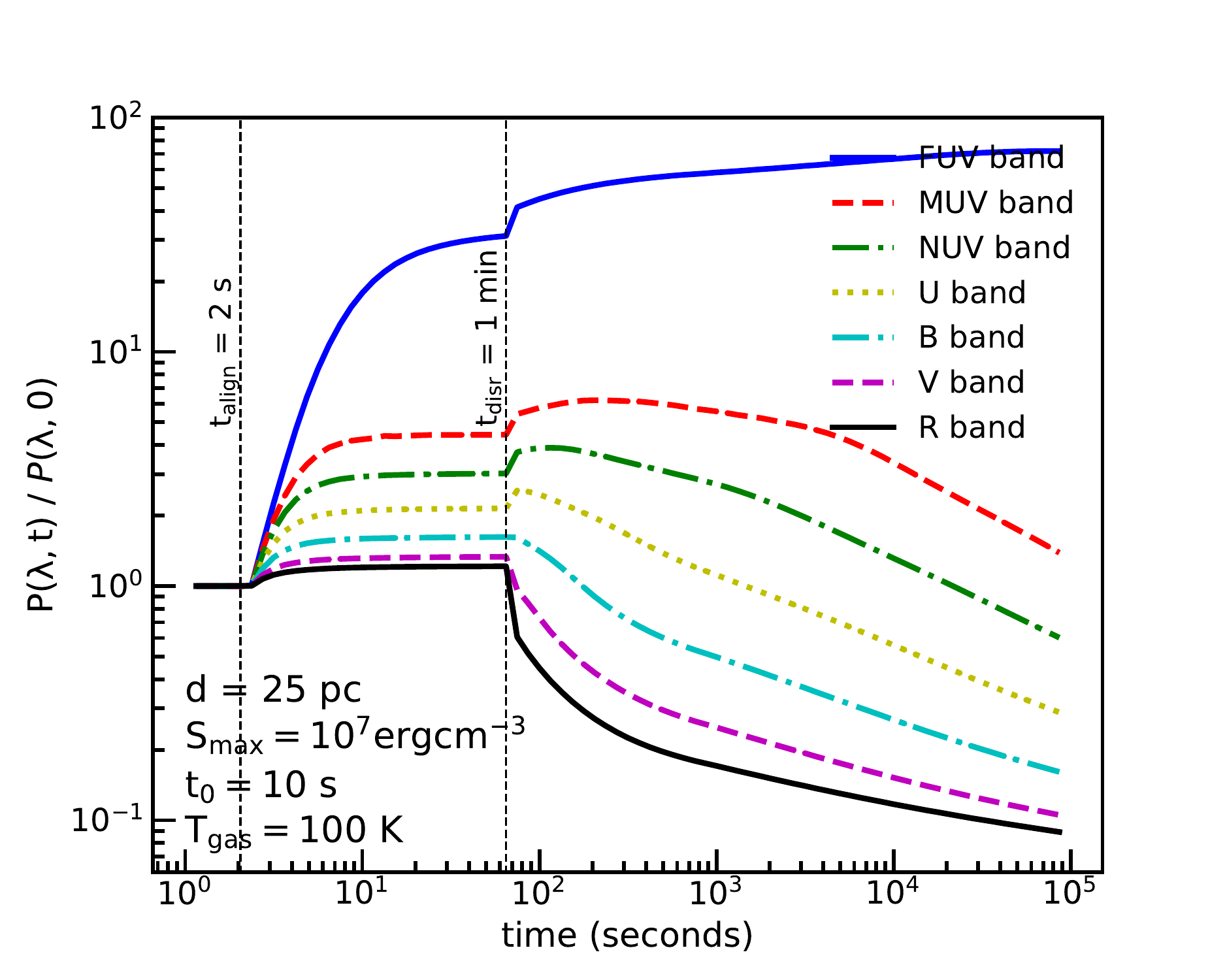}
\includegraphics[width=0.45\textwidth]{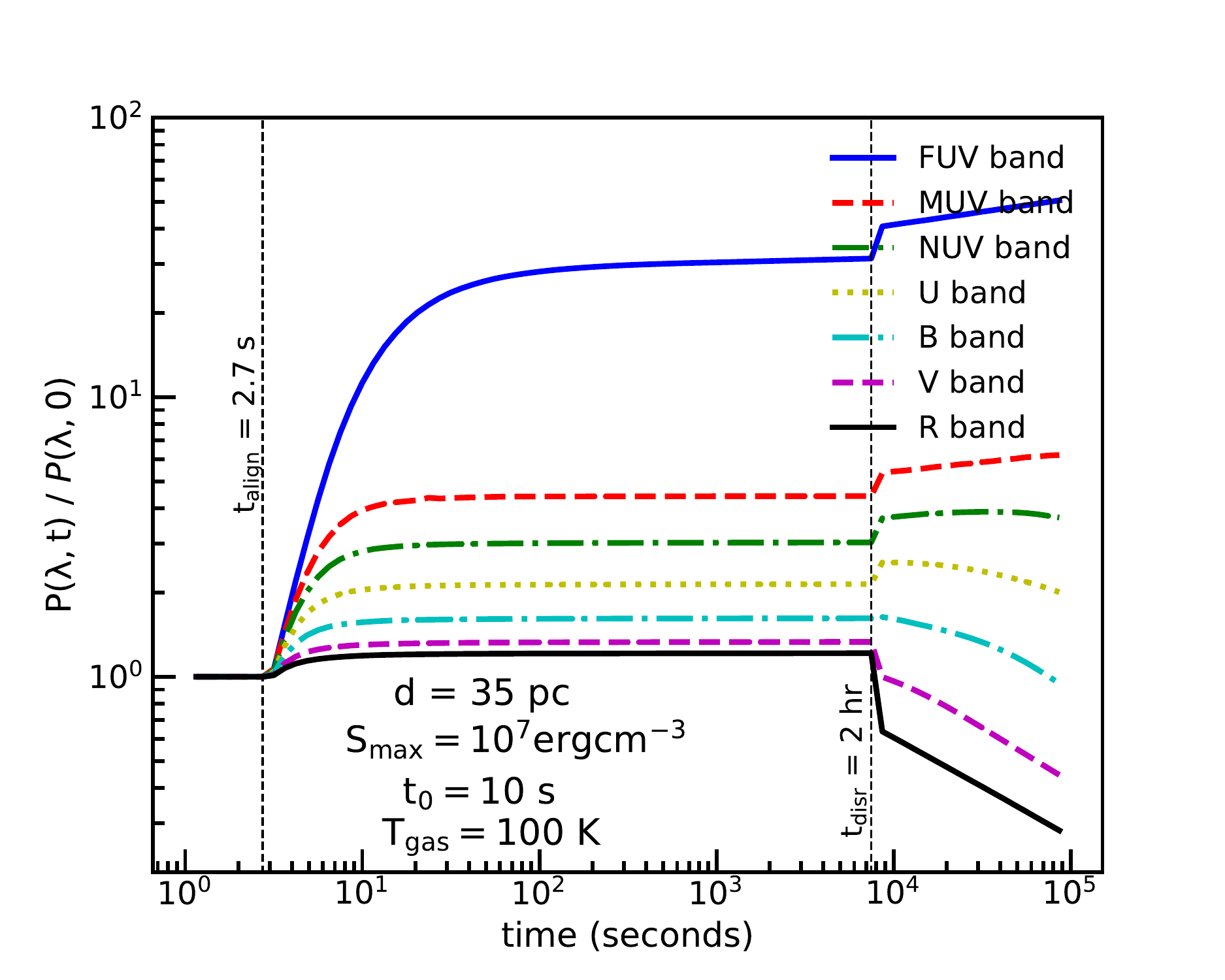}
\caption{Same as Figure \ref{fig:Aext_dtime} but for dust polarization by aligned grains, assuming $T_{\rm gas}=100$ K. $P(\lambda)$ first changes since alignment time $t_{\rm align}$ when the radiation field from GRB afterglows dominate in grain alignment then second changes after the disruption time $t_{\rm disr}$ due to the removal of large grains. At 15 pc, the polarization in U bands first increases up to 20 s due to increasing grain alignment by RATs then decreases rapidly due to disruption of large grains. In contrast, the polarization from FUV to NUV continues to rise.}
\label{fig:Polabs_dtime}
\end{figure}

Figure \ref{fig:Polabs_dtime} shows the time variation of the polarization degree of GRB afterglows, $P(\lambda,t)/P(\lambda,0)$, evaluated in the different bands for three dust cloud distances, assuming $T_{\rm gas}=100$ K. After the alignment time, the polarization degree in all bands increases significantly due to the decrease of $a_{\rm align}$ as a result of the enhanced radiation field, then mostly saturates after about $t\lesssim 30$ s. After that, the disruption happens (see Figure \ref{eq:adisr}, lower panel) that makes the optical/NIR polarization degree start to decline rapidly, but the UV polarization continues to rise due to the enhancement of small aligned grains and can decrease slightly later if its suitable grain size is removed, i.e., the case of UV polarization degree given by grains at 15 pc. All variations will stabilize when the grain disruption size reaches its saturated value after a long time. A distant cloud makes the polarization curve change its phase later, i.e., longer $t_{\rm align}$ and $t_{\rm disr}$, and vary $P(\lambda)$ in the small range than a nearby cloud. For example, the phase that grains are only aligned by RATs lasts from $t_{\rm align}=13$ s for clouds at 15 pc to near one minute and two hours for clouds at 25 pc and 35 pc. Besides, after one day, the polarization degree in R band only decreases $3-10$ times for grains at 25 pc and 35 pc, but it is $\sim$ 25 times for grains at 15 pc.
 
%\subsection{Time-variability of the maximum polarization wavelength}

\begin{figure}
\includegraphics[width=0.5\textwidth]{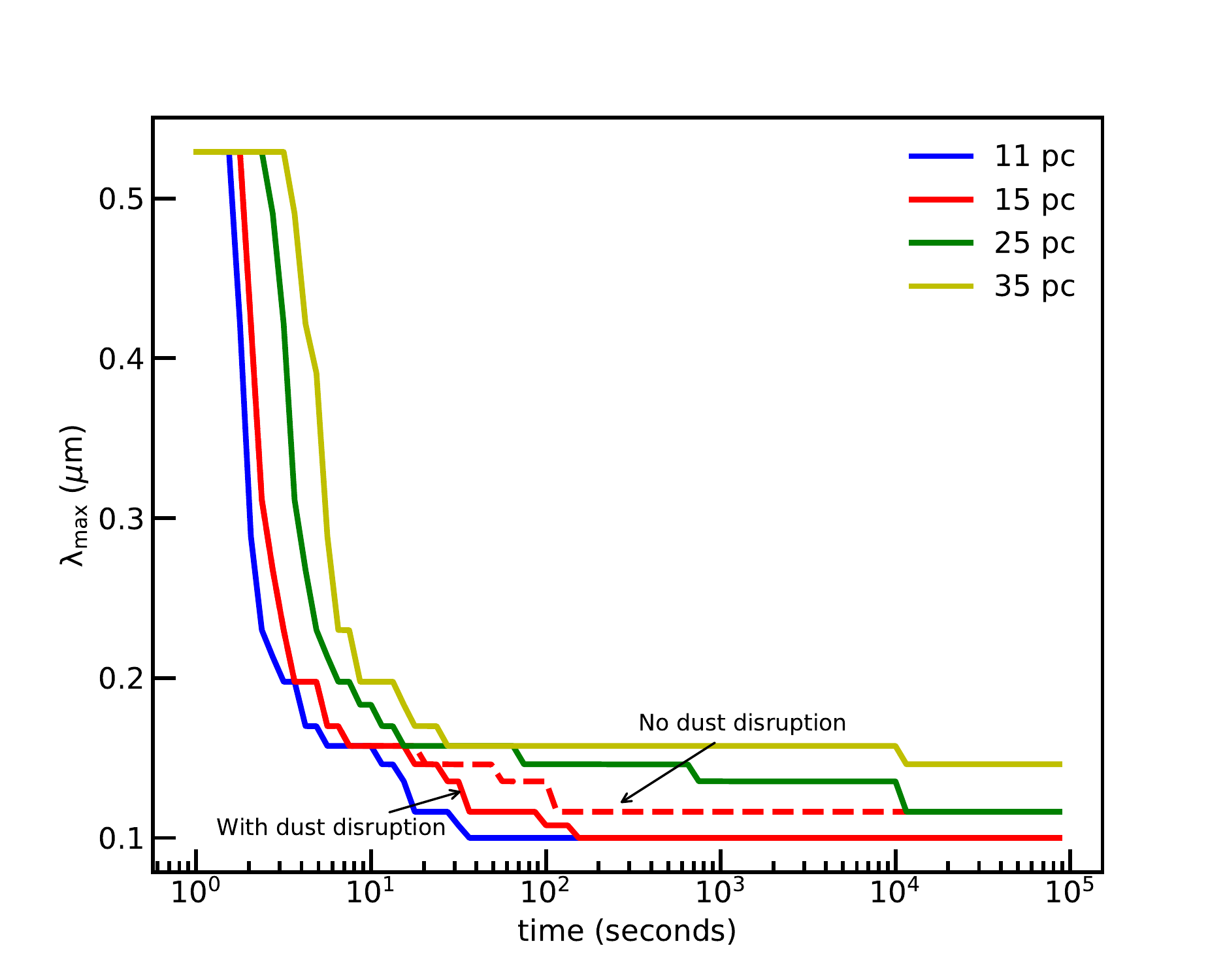}
\caption{Time-variation of maximum polarization wavelength $\lambda_{\rm max}$ for grains at distance from 11 pc to 35 pc with RATD (solid line) and without RATD (dashed line), assuming $T_{\rm gas}=100$ K.}
\label{fig:lambda_time} 
\end{figure}

Figure \ref{fig:lambda_time} shows the variation of the maximum polarization wavelength $\lambda_{\rm max}$ during one day for grains at different distances and the comparison of $\lambda_{\rm max}$ with (solid line) and without (dashed line) the grain disruption process for clouds at 15 pc. The peak wavelength remains constant at $\lambda_{\rm max}=0.55\mum$ due to grains aligned by an average interstellar radiation field from $t=0$ s to $t_{\rm align}$. Beyond that, it decreases suddenly due to the enhanced alignment of small grains (i.e., smaller $a_{\rm align}$; see Figure \ref{fig:align_time}) and continues to change slowly when the disruption begins, i.e., the solid and dashed line of the $\lambda_{\rm max}(t)$ line for cloud at 15 pc (see Figure \ref{fig:Polabs_time}). For a distant cloud, $\lambda_{\rm max}$ decreases later and gives a higher value than ones given by a nearby cloud due to a weaker radiation strength (Figure \ref{fig:Polabs_dtime}). 

\section{Effect of RATD on the light curves of GRB afterglows}\label{sec:lightcurve}
As shown in Section \ref{sec:GRBext}, RATD increases dust extinction in FUV-NUV bands but decreases dust extinction in optical-NIR bands due to the conversion of large grains into smaller ones. Such a variation of dust extinction by RATD would change the observed spectrum of GRB afterglows as well as their light curves. In this section, we apply the new extinction curves in the presence of RATD to study how it affects the observed light curve of GRB afterglows. 
 
Let $\tau(\lambda,t) = A(\lambda,t)/1.086$ be the optical depth induced by dust extinction from an intervening cloud between the GRB and an observer which is measured at time $t$ since the burst. The specific luminosity of GRB afterglows observed at time $t$ on the Earth is given by
\bea\label{eq:Lumi}
L_{\lambda}(t) = L_{\lambda}(0) e^{-\tau(\lambda,t)},
\ena
where $L_{\lambda}(0)$ is the intrinsic specific luminosity given by Equation (\ref{eq:Lnu_RS}) and (\ref{eq:Lnu_FS}).

For our calculations, we assume that the intervening cloud has a visual extinction of $A_{\rm V}=2$ mag at $t=0$ s, which corresponds a total gas column density of $N_{\rm H}=3.14\times 10^{21}\cm^{-2}$. The choice of $A_{\rm V}=2$ is intended to reflect a dusty environment surrounding GRBs. For the given $N_{\H}$, one can calculate $\tau(\rm \lambda)$ using $A(\lambda,t)$ calculated in Section (\ref{sec:extintion}), and the observed luminosity $L_{\rm \lambda}(t)$ is calculated via Equation (\ref{eq:Lumi}).

\begin{figure}[htb!]
      \includegraphics[width=0.45\textwidth]{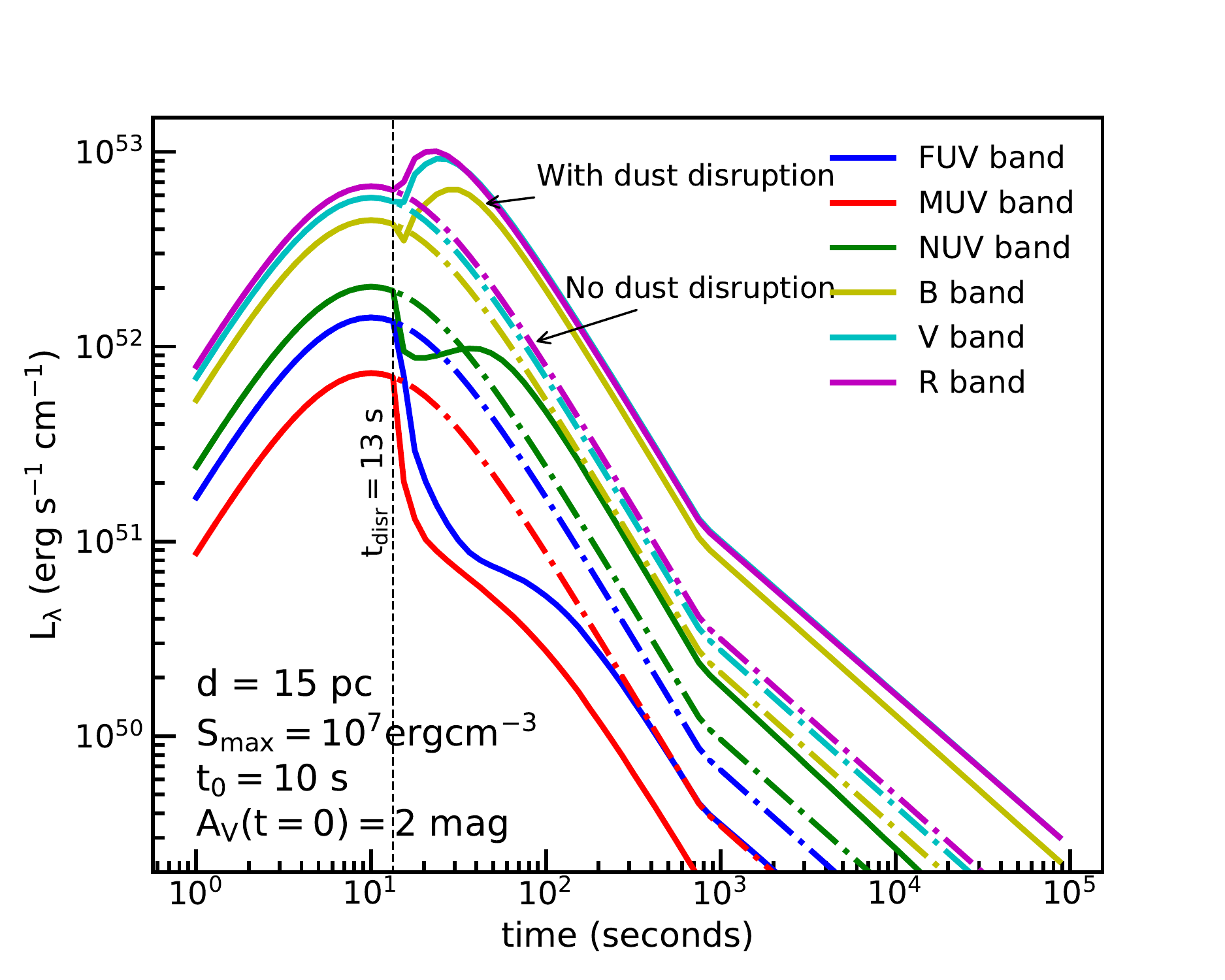}
      \includegraphics[width=0.45\textwidth]{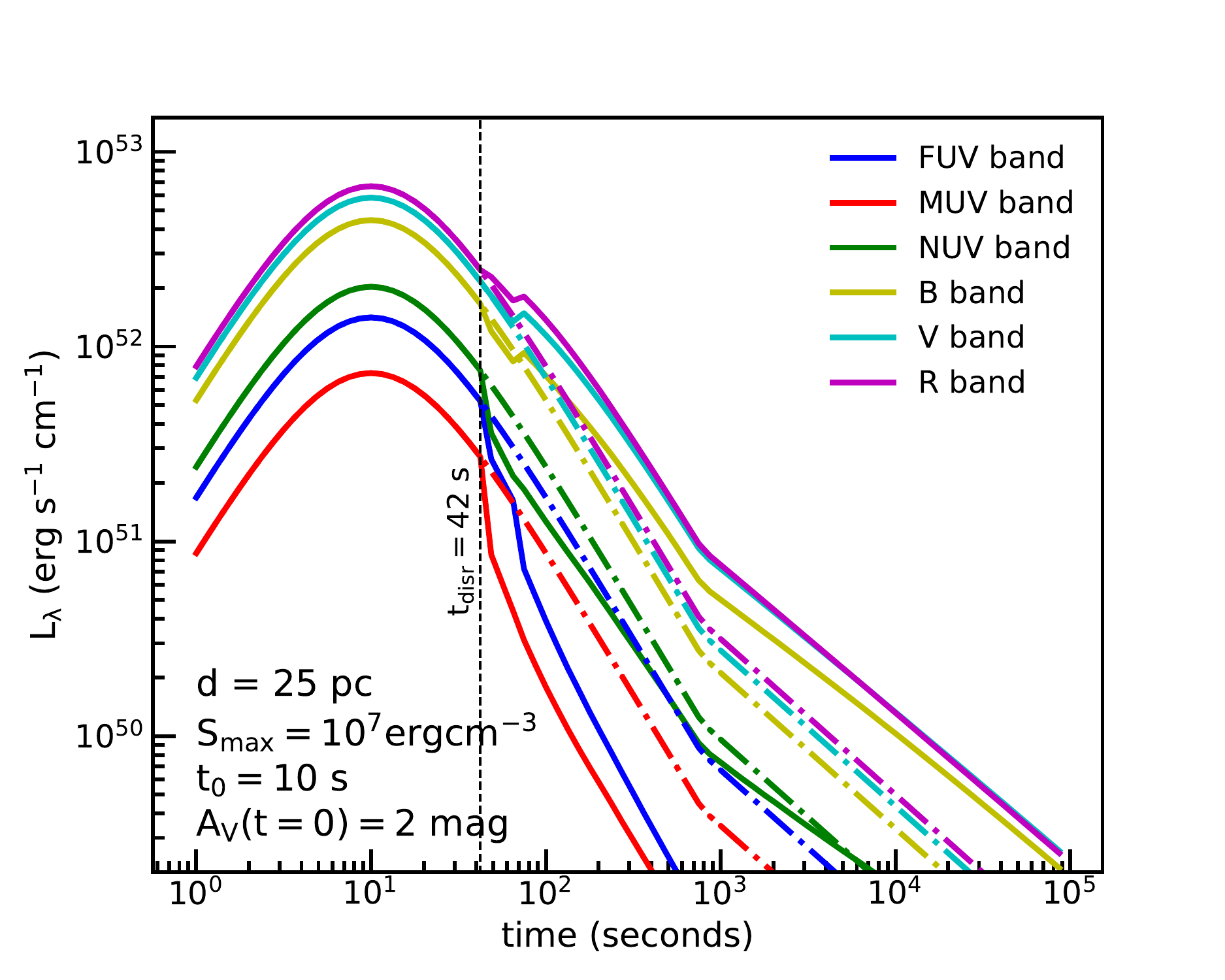}
      \includegraphics[width=0.45\textwidth]{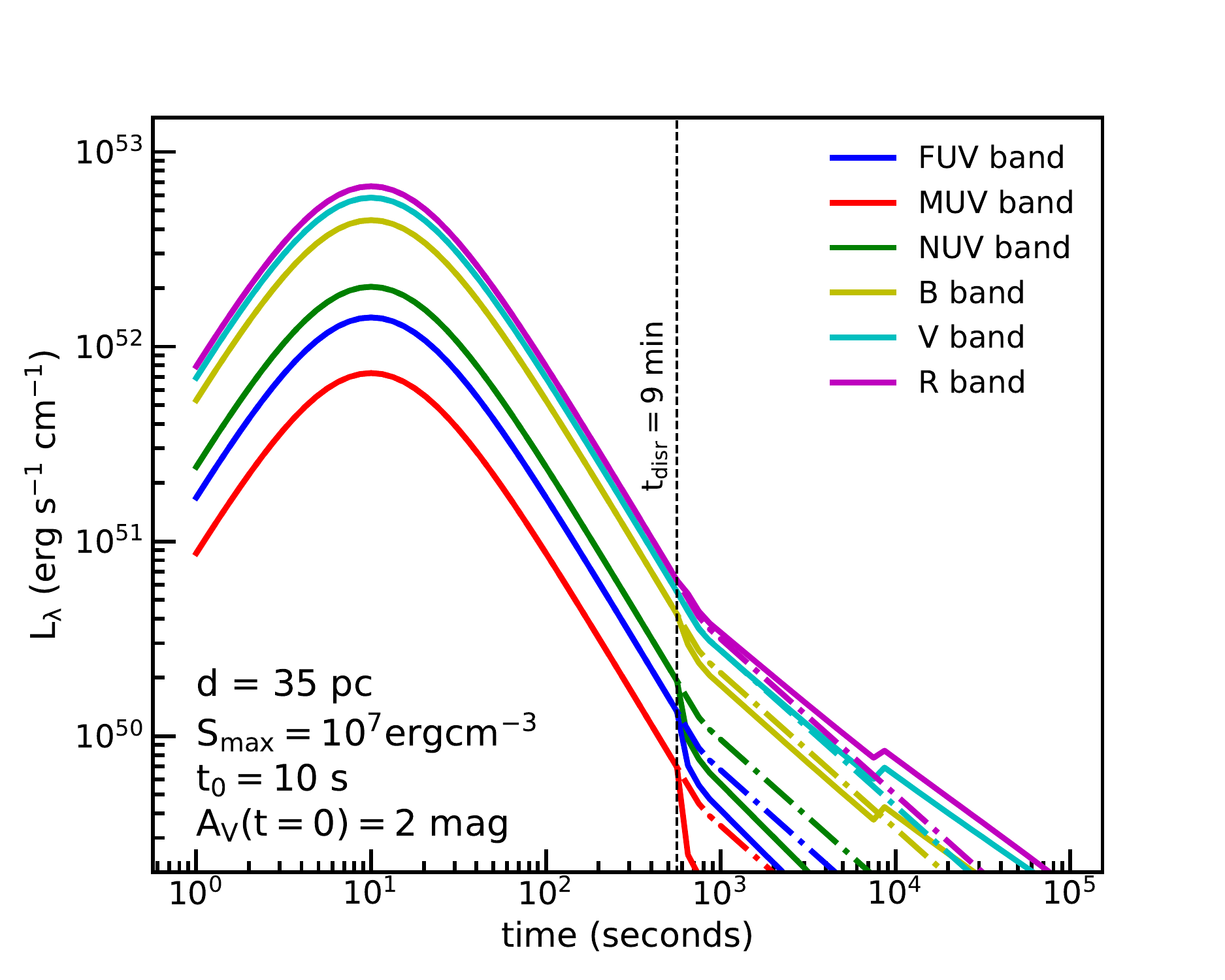}
\caption{Time-dependence of the specific luminosity $L_{\rm \lambda}(t)$ from FUV to NIR band for a dust cloud located at 15 pc, 25 pc, and 35 pc with (solid line) and without (dashed line) RATD. The optical extinction $A_{\rm V}=2$ mag at $t=0$ s is adopted to calculate the column density $N_{\rm H}$. GRB afterglows appear brighter in the visible band and dimmer in the FUV-NUV bands when large grains are destroyed by RATD.}
\label{fig:Iv_dtime}
\end{figure}

Figure \ref{fig:Iv_dtime} shows the time-variation of the observed light curve from FUV to R bands after entering a dust cloud at 15 pc, 25 pc, and 35 pc, with (solid line) and without (dashed line) grain disruption.\footnote{For a dust cloud of thickness $1$ pc, the volume density is $n_{\rm H}=N_{\rm H}/1\rm pc\sim 10^{3}\cm^{-3}$, i.e., the cloud is dense.}

One can see that after the disruption time of $t_{\rm disr}=13$ s for clouds at $d=15$ pc, GRB afterglows suddenly become 're-brightening' up to $\sim$ 3 times in the visible-NIR bands compared with no grain disruption case. The reason is that the reduction of the visible-NIR extinction will let more light escape from dust, resulted in the increase of visible-NIR luminosity. In contrast, the increase of the UV extinction will block more short-wavelength photons, that makes GRB afterglows become 'dimmer' from 3-5 times in the FUV-NUV band compared with the case of no dust disruption (Figure \ref{fig:Iv_dtime}, upper panel). 

When the cloud distance increases, these features will happen later and exhibit the smaller amplitude than ones given by nearby clouds (Figure \ref{fig:Iv_dtime}, central and lower panel). Besides, at nearby clouds, i.e., $d\leq 15$ pc, the luminosity in FUV and MUV bands can increase slightly after a long time compared with before ($\sim$ 100 s) due to the disruption of small grains, while it does not happen with distant clouds, i.e., $d=$ 25 pc ad 35 pc. In addition, one may not obtain any change in the observed light curve if clouds locate very far from GRB afterglows, where RATD can not destroy grains effectively, i.e., $d> 40$ pc
 
\section{Discussion}\label{sec:discuss}
\subsection{Comparison of RATD to thermal sublimation and Coulomb explosion}
GRBs are expected to explode in a dusty region (\citealt{Morgan14}), such that intense radiation field from GRBs can have important effects on the surrounding environment. This in turn affects the observed light curves and color of GRB afterglows. Therefore, dust destruction by GRBs was studied extensively in literature.

\cite{Waxman20} first studied sublimation of dust grains by prompt optical-UV emission of GRBs and found that dust grains up to $\sim 10$ pc can be completely evaporated. Later, \cite{Fruchter01} studied dust destruction caused by X-ray irradiation and found that grains can be disrupted by X-ray heating and charging (i.e., Coulomb explosions) to distances of $\sim 10$ and $\sim 100$ pc, respectively. The effective timescales of both sublimation and Coulomb explosions is short, $t\lesssim 10-100$ s, after the start of the burst. However, the issue of photoelectric yield by X-ray charging is not studied in detail in \cite{Fruchter01}. As shown in \citealt{Hoang15b}, the yield for large grains of $a\sim 1\mum$ is one order of magnitude lower than that of $a\sim 0.001\mum$. Thus, similar to grain sublimation, Coulomb explosions are more efficient for small grains because those grains have higher photoelectric yield and a lower critical charge for explosions (\citealt{Hoang15b}).

In this paper, we study rotational disruption of dust induced by irradiation of optical-UV GRB afterglows using the Radiative Torque Disruption (RATD) mechanism. We find that grains can be disrupted up to distances of about 40 pc, on a timescale up to days, much longer than sublimation and Coulomb explosions which rely on the prompt emission phase. The disruption time depends on grain sizes, the maximum tensile strength, and the distance to the source (see Figures \ref{fig:adisr_time} and \ref{fig:adisr_Smax}). 

One of the key differences between RATD and thermal sublimation and Coulomb explosions is that RATD increases the abundance of small and very small grains relative to large ones, keeping the total dust mass constant. As a result, optical-NIR extinction decreases, but UV extinction increases with time (see Figure \ref{fig:Aext_time}). On the other hand, sublimation is more efficient for small grains and transforms dust to gas, such that dust extinction at all wavelengths and color excess decrease with time (\citealt{Perna02}; \citealt{Perna03}). 

Both thermal sublimation and Coulomb explosions by X-rays can significantly change dust properties during the prompt emission phase of GRBs of $t\lesssim 10-100$ s after the burst. As a result, very early phase observations are required to test time-variation of dust extinction and polarization by these mechanisms (\citealt{Perna03}). In contrast, RATD relies on optical GRB afterglows that lasts on longer timescales of days. Therefore, observational testing of RATD appears to be much easier.

\subsection{Predictions of Observational Properties for GRB afterglows induced by RAT alignment and RATD}
Below we summarize four main predictions for observational properties of GRB afterglows induced by an intervening dust cloud when the effects of grain alignment and disruption by intense GRB afterglows are taken into account.

\subsubsection{Prediction 1: RATD decreases Optical-NIR Extinction and $R_{\rm V}$ over time}
In Section \ref{sec:extintion}, we have shown that RATD can destroy large grains around GRB afterglows up to 40 pc for an optically thin environment. The depletion of large grains by RATD decreases the optical-NIR extinction but increases the UV extinction. Moreover, we predict that the values of $R_{\rm V}$ gradually decrease from the standard value of $R_{\rm V}=3.1$ to $R_{\rm V}\sim 1$ in the presence of RATD. Therefore, the extinction curves toward GRB afterglows that have a dust cloud nearby would be different from the standard Milky Way (MW) extinction curve, which exhibits a steep far-UV rise due to high abundance of small grains (see Figure \ref{fig:Aext_time}).

\subsubsection{Prediction 2: RATD increases and then decreases the color excess of GRB afterglows}
Our theoretical results from Figure \ref{fig:EBV} predict that the color excess $E(B-V)$ changes with time. It first increases rapidly and then decreases with time after the peak. The peak of $E(B-V)$ depends on the cloud distance and grain properties.

\subsubsection{Prediction 3: RATD increases and then decreases Optical-NIR polarization}
Subject to an intense radiation of GRB afterglows, dust polarization first rises quickly due to enhanced alignment of small grains by RATs. At the same time, the peak wavelength $\lambda_{\rm max}$ shifts to smaller wavelengths. This process continues from $t_{\rm align}$ to $t_{\rm disr}$. When RATD begins, the optical-NIR polarization decreases substantially due to the depletion of large grains, whereas UV polarization increases due to the increased abundance of small grains (see Figure \ref{fig:Polabs_time}). The exact values of $t_{\rm align}$ and $t_{\rm disr}$ depend on the radiation field, dust properties, and distance of dust clouds to the source. 

\subsubsection{Prediction 4: RATD produces an optical-NIR re-brightening of GRB afterglows}
Due to the decrease of optical-NIR extinction, the observed flux of GRB afterglows in optical-NIR bands is spontaneously increased after disruption time (see Figure \ref{fig:Iv_dtime}). The RATD effect induces the re-brightening in optical-NIR bands, which occurs at disruption time $t_{\rm disr}$. The re-brightening time depends on the cloud distance to the source and dust properties (e.g., tensile strength) as shown in Figure \ref{fig:adisr_time}). 

\subsection{Comparison of observed properties of GRB afterglows with model predictions}

First, observations of GRB 120119A by \cite{Morgan14} show a decrease of visual extinction from $A_{\rm V}\sim 1.55$ at $t\sim 10$ s to $A_{V}\sim 1.1$ at $t\sim 100$ s after the burst, corresponding to a decrease of
$30\%$ over a period of 10-100 s. Such a rate of the decrease is several times larger than theoretical predictions for the $t\sim 10-100$ s period using dust sublimation induced by prompt emission because sublimation is most efficient for $t<10$ s (see Fig. 5 in \citealt{Perna03}). However, this fast decrease in $A_{V}$ is consistent with our first prediction (see e.g., Figure \ref{fig:Aext_time}). Moreover, photometric observations of GRB afterglows show that a SMC-like extinction curve with a steep far-UV rise is preferred for GRBs (\citealt{Schady12}; \citealt{Schady17} for a review; \citealt{Heintz:2017}). \cite{2018A&A...609A..62B} also found that the extinction toward GRBs at redshifts $z>4$ are best-fitted with a SMC-like extinction curve. In particular, previous studies (e.g., \citealt{Zafar:2018}; \citealt{Zafar:2019}) show that the majority of light of sight toward GRB afterglows have lower values of $R_{V}< 3.1$ (see Table 2 in \citealt{Zafar:2018}). The observed features mentioned above require an increased abundance of small grains from the standard interstellar dust model (e.g., \citealt{Schady10}). The conversion of large grains into smaller ones via RATD is the plausible mechanism to explain this feature (i.e., our first prediction). 

\begin{figure}
\includegraphics[width=0.5\textwidth]{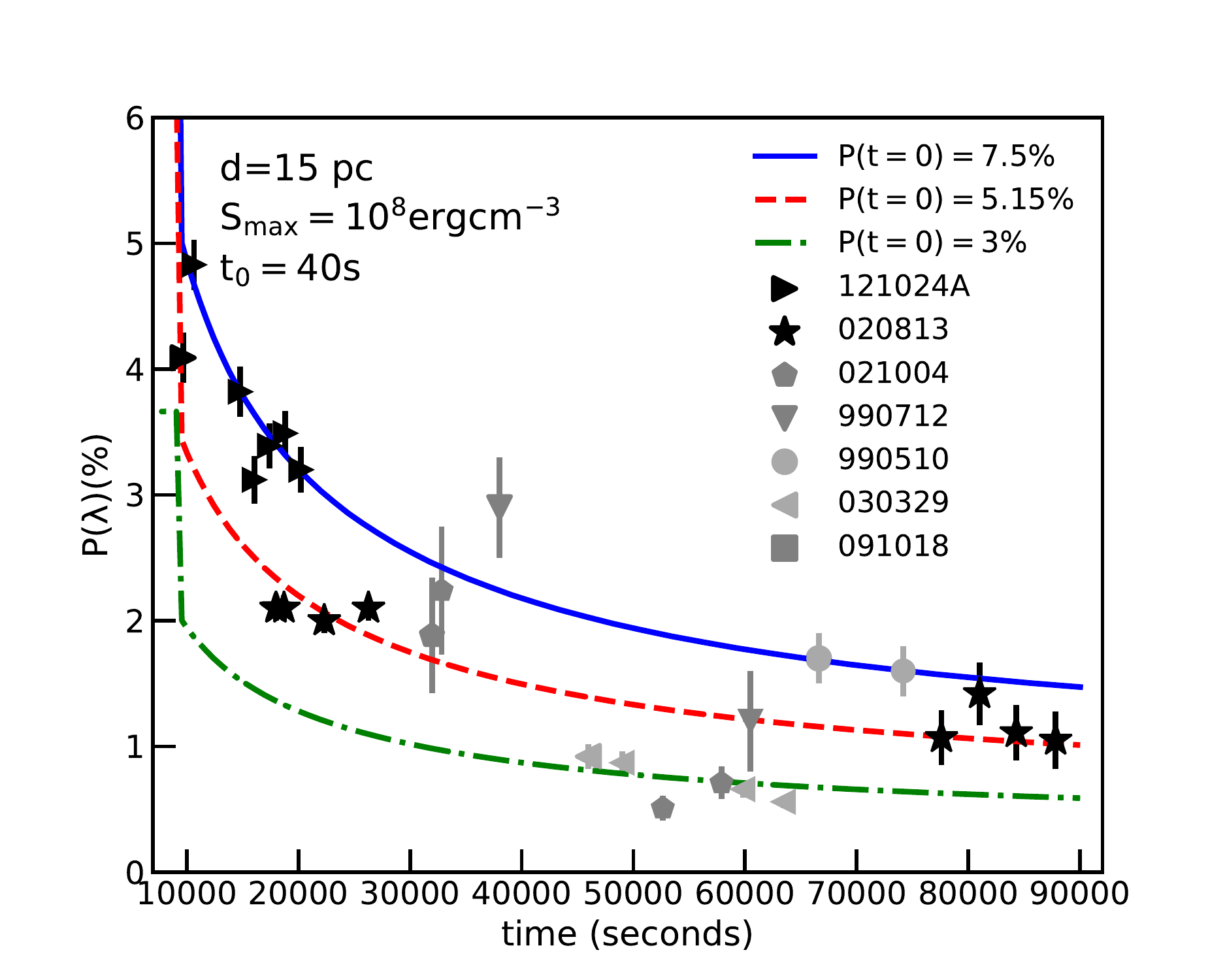}
\caption{Time-variation of optical polarization of GRB afterglows (see \citealt{Covino:2016vo} for details) compared with our theoretical model in the R band, assuming a dust cloud at 15 pc and tensile strength $S_{\rm max}=10^{8}\erg\cm^{-3}$. The original polarization degree in the R band $P(t=0)$ is varied to fit the observational data. }
\label{fig:P_observe}
\end{figure}

Second, photometric observations of GRB afterglows usually show a significant red-to-blue colour change after the trigger (see e.g., \citealt{Nardi14}), which is partly suggested to be a result of photodestruction of surrounding grains (\citealt{Morgan14}). However, the popular mechanisms of dust destruction cannot support this scenario due to the inconsistency between its timescale and the observed time. For example, \cite{Morgan14} reported a significant red-to-blue color during 200 s after the burst toward GRB 120119A, and similar effect is reported by \cite{Perley10} for GRB 061126, which are longer than predicted by previous dust destruction mechanisms. The observed feature is however consistent with our second prediction by RATD. As shown in Figure \ref{fig:EBV}, our model of the time-variation of color excess $E(B-V)$ for $S_{\rm max}=10^{8}\erg\cm^{-3}$ and $d=15\sim 20$ pc can reproduce well their observational timescale.  

%This observed feature is however along with our second prediction by RATD.

Third, polarimetric observations usually report time-variability of optical polarization of GRB afterglows on a timescale of hundred seconds to days (see \citealt{Covino:2016vo} for a review). Such a long timescale variability is difficult to reconcile with thermal sublimation by prompt emission, but consistent with our third prediction. 
Indeed, in Figure \ref{fig:P_observe}, we compare the optical polarization observed toward various GRBs with our theoretical models. For  GRB 020813 (star symbols), \cite{Barth03} showed that the optical polarization degree of decreases from $2.4-1.8\%$ during $4.7-7.9$ hr, and subsequent observations after two days by \cite{Covino03a} give a much lower level of $p=0.8\pm 0.16\%$ with the stable polarization angle. This data are in good agreement with our model, where the original polarization of dust is $P(V,t=0)=7.5\%$. Similarly, for GRB 990712 (triangle symbols), \cite{Rol20} also report the variation of optical polarization degree from $2.9 \pm 0.4\%$ after 0.44 day to $1.2\pm0.4\%$ after 0.7 day and $2.2\pm 0.7\%$ after 1.45 days after the burst. For GRB 021004, \cite{Covino16} shows the optical band decreasing from $1.88-0.71\%$ within one day. Furthermore, \cite{Covi03} find a steep decrease in the polarization degree of GRB 030329, from $0.9\pm0.1\%$ in the blue light to $0.5\pm0.1\%$ in the red light after 3.6 h after the explosion. 

%Recently, \cite{Jordana19} observed that variation in the polarization in BVRI bands of GRB 190114C. \textbf{It shows that the polarization in BV band tends to increase from $2\%$ to $4.5\%$ during 200-2000s, while decreases from $3.8\%$ to $2.1\%$ in R band after the same time. This decreasing of optical polarization are consistent with our second prediction. The time-varying optical polarization degree of these above GRB afterglows are shown in Figure \ref{fig:P_observe}. To check that if our model can reproduce the observational data, we assume a dust cloud at 15 pc and find that the model that grains have $S_{\rm max}=10^{9}\erg\cm^{-3}$ and GRB afterglows peak at $t_{0}=40$ s can explain for these variations of $P(\%)$ in R band}.
 
\subsection{Origins of optical re-brightening of GRB afterglows}

Late-time observations of GRB afterglows frequently report a re-brightening in their optical-NIR light curves. For instance, \cite{Klotz2005} detected a re-brightening at about 33 min from the GRB 050515a afterglow. Using the data from Gamma-Ray burst Optical Near-infrared Detector (GROND) on board of {\it SWIFT} satellite, \cite{Nardi11} found a fast optical re-brightening of GRB 081029 around 0.8 hr after the burst, and \cite{Greiner13} detected a re-brightening for GRB 100621A at 1 hr. Moreover, \cite{Nardi14} found the re-brightening of GRB 100814A after 0.3 days, and \cite{Kann18} found the re-brightening of GRB 111209A at 0.8 days. Recently, \cite{ugarte18} found a rapid optical re-brightening at 2.4 hr from GRB 100418A. 

\begin{figure}
\includegraphics[width=0.5\textwidth]{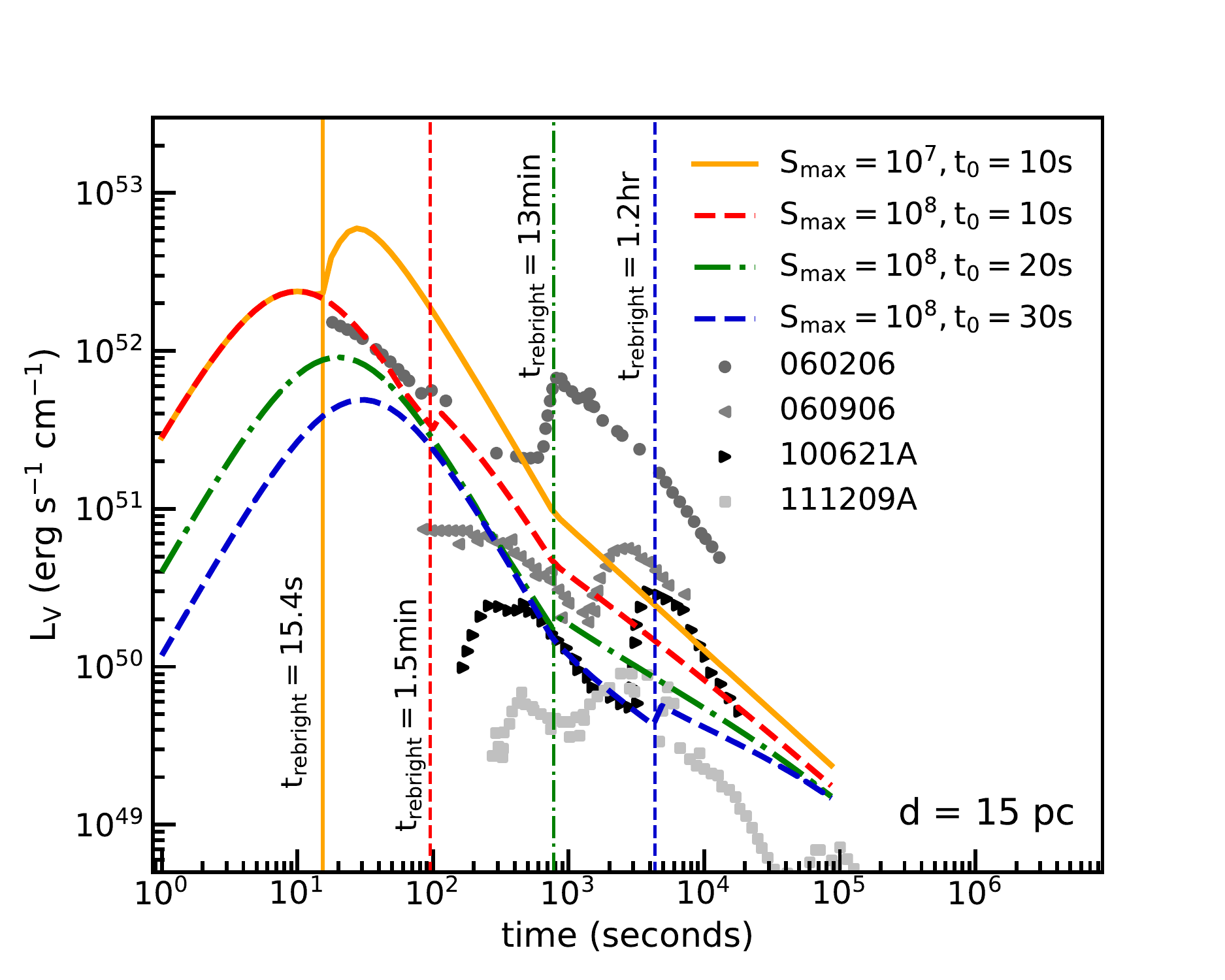}
\caption{Comparison of our predicted light curves with the observed light curves in R band for GRB afterglows with a prominent optical re-brightening (see \citealt{ugarte18} for GRB data), assuming a dust cloud of $A_{V}(t=0)=3$ at 15 pc from the source and different model parameters ($S_{\rm max},t_{0})$). The optical re-brightening time can be reproduced by the theoretical models.}
\label{fig:rebright}
\end{figure}

The nature of such an optical re-brightening is unclear. Several processes were proposed to explain this feature, including intrinsic processes related to the central engine, external shocks, and dust extinction effect (see \citealt{Nardi14} for details). To study whether our models can reproduce the optical re-brightening, in Figure \ref{fig:rebright}, we plot the light curves of four GRB afterglows with an optical re-brightening (\citealt{ugarte18}) and compare with our theoretical predictions with two model parameters ($S_{\rm max}$ and $t_{0}$). Our models for a dust cloud of original visual extinction $A_{V}(t=0)=3$ can indeed reproduce the timing of optical re-brightening, although the models yield a lower amplitude of the re-brightening than observations. Increasing the original extinction $A_{V}(t=0)$ can increase the re-brightening amplitude and better fits the observational data. Note that the contribution of other mechanisms (e.g., central engine and external shocks) cannot be ruled out as a cause of the optical re-brightening.

\subsection{Effect of light attennuation by intervening dust on RATD}\label{sec:reddening}

So far, we have considered grain rotational disruption by GRB afterglows by disregarding the effect of intervening dust. In this case, RATD can disrupt grains up to 40 pc just after about one day. In realistic situations, intervening grains will attenuate the GRB radiation field, which will reduce the efficiency of RATD.
 
We assume that the GRB afterglow emits from the center of a dusty bubble that has a central cavity of radius $10$ pc, which is presumably cleared out by sublimation (\citealt{Waxman20}) or Coulomb explosions (\citealt{Fruchter01}) during the prompt emission. To calculate the grain disruption size in the presence of dust reddening, we divide the intervening cloud into slabs of the same thickness $\Delta d$ and assume that grains in slice $n$ are disrupted to the same value of $a_{\rm disr,n}$. Let $\tau_{\rm n}=A(\lambda, n)/1.086$ be the optical depth induced by dust grains in the $nth$ slice (see Equation \ref{eq:Aext}). The radiation energy density $u_{\rm rad,n}$ at slice $n$ is given by
\bea \label{eq:urad}
u_{\rm rad,n} = \frac{L_{\rm bol} \times e^{-\tau_{n}}}{4 \pi c d^{2}},
\ena
where $\tau_{n}$ is the total effective optical depth given by grains from $d= 10$ pc to the slice $n$, which is equal to:
\bea
\tau_{n} = \sum_{i=0}^{n-1} \tau_{i},
\ena
with $\tau_{i}$ being the effective optical depth at slice $i$ defined as $e^{-\tau_{i}}=\int_{1eV}^{13.6eV} u_{\rm \lambda,i} e^{-\tau_{\lambda,i}} d\lambda/ u_{\rm rad,i}$ (\citealt{Hoang19}).

\begin{figure}[htb!]
\includegraphics[width=0.5\textwidth]{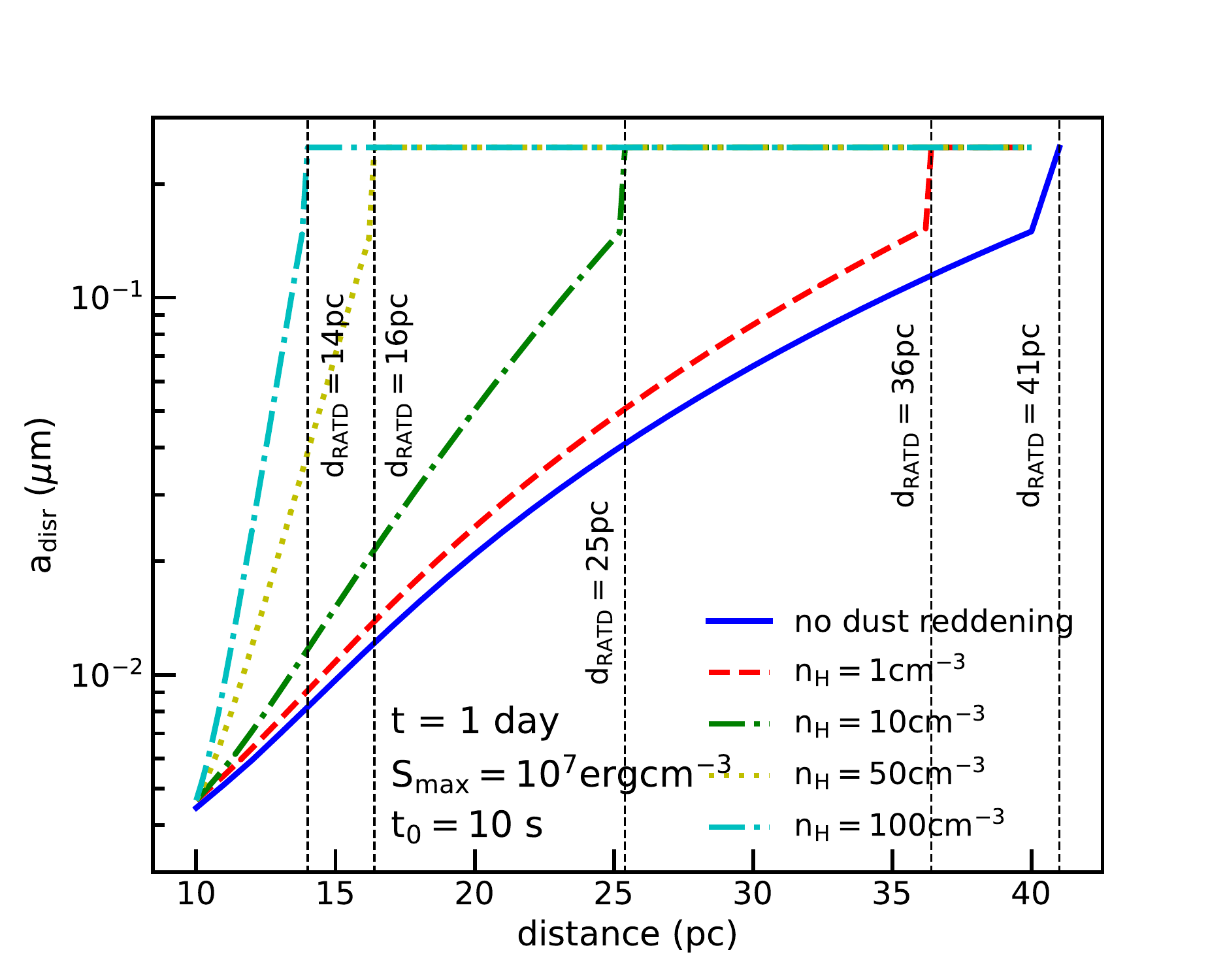}
\includegraphics[width=0.5\textwidth]{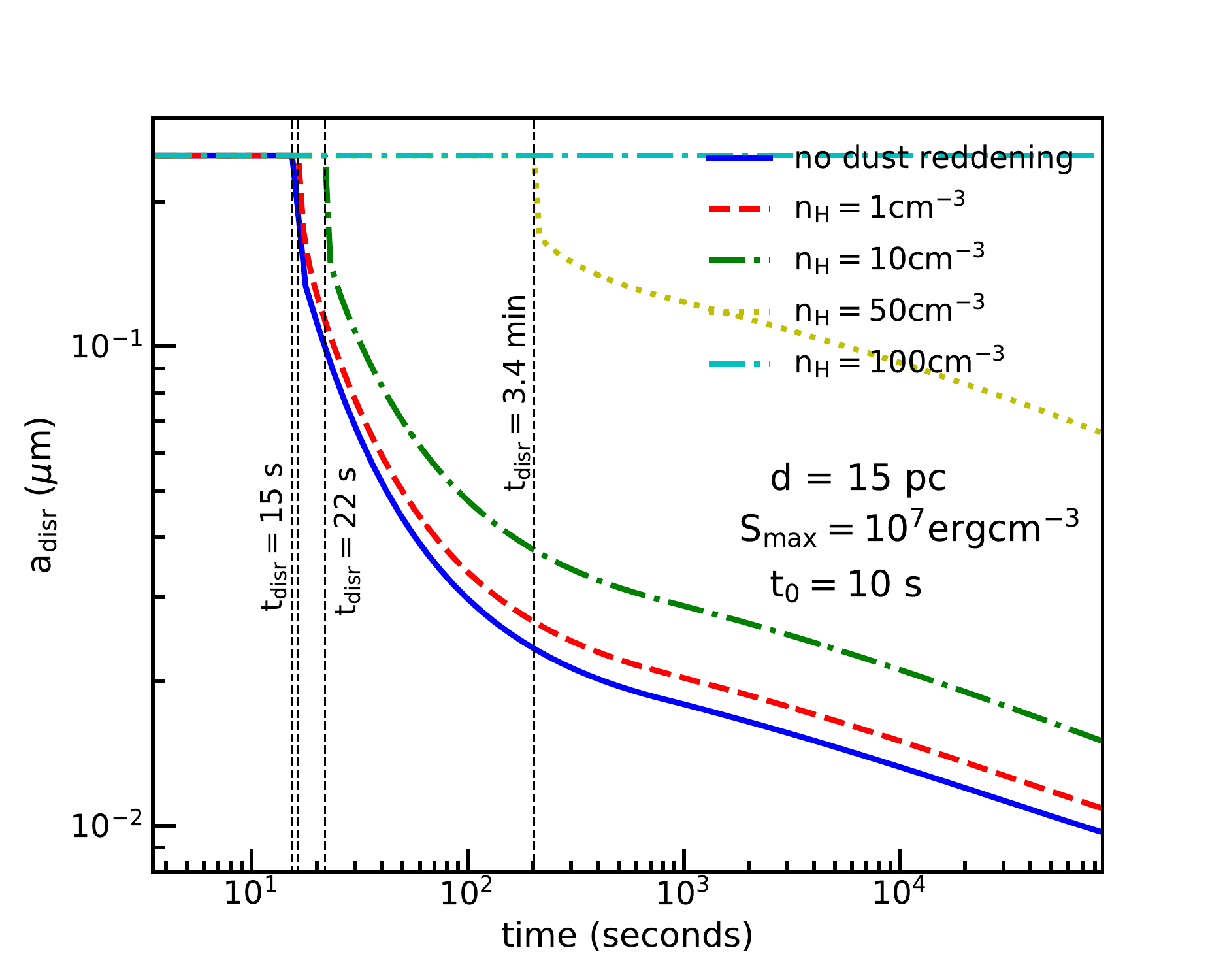}
\caption{Comparison of disruption size with and without light attenuation by intervening dust. The upper panel shows $a_{\rm disr}$ vs. distance after one day, and the lower panel shows $a_{\rm disr}$ vs. time for a cloud at 15 pc, assuming $S_{\rm max}=10^{7}\erg\cm^{-3}$ and different gas density from $1-100\cm^{-3}$.}
\label{fig:disrGRB_nH}
\end{figure}

\begin{figure}[htb!]
\includegraphics[width=0.5\textwidth]{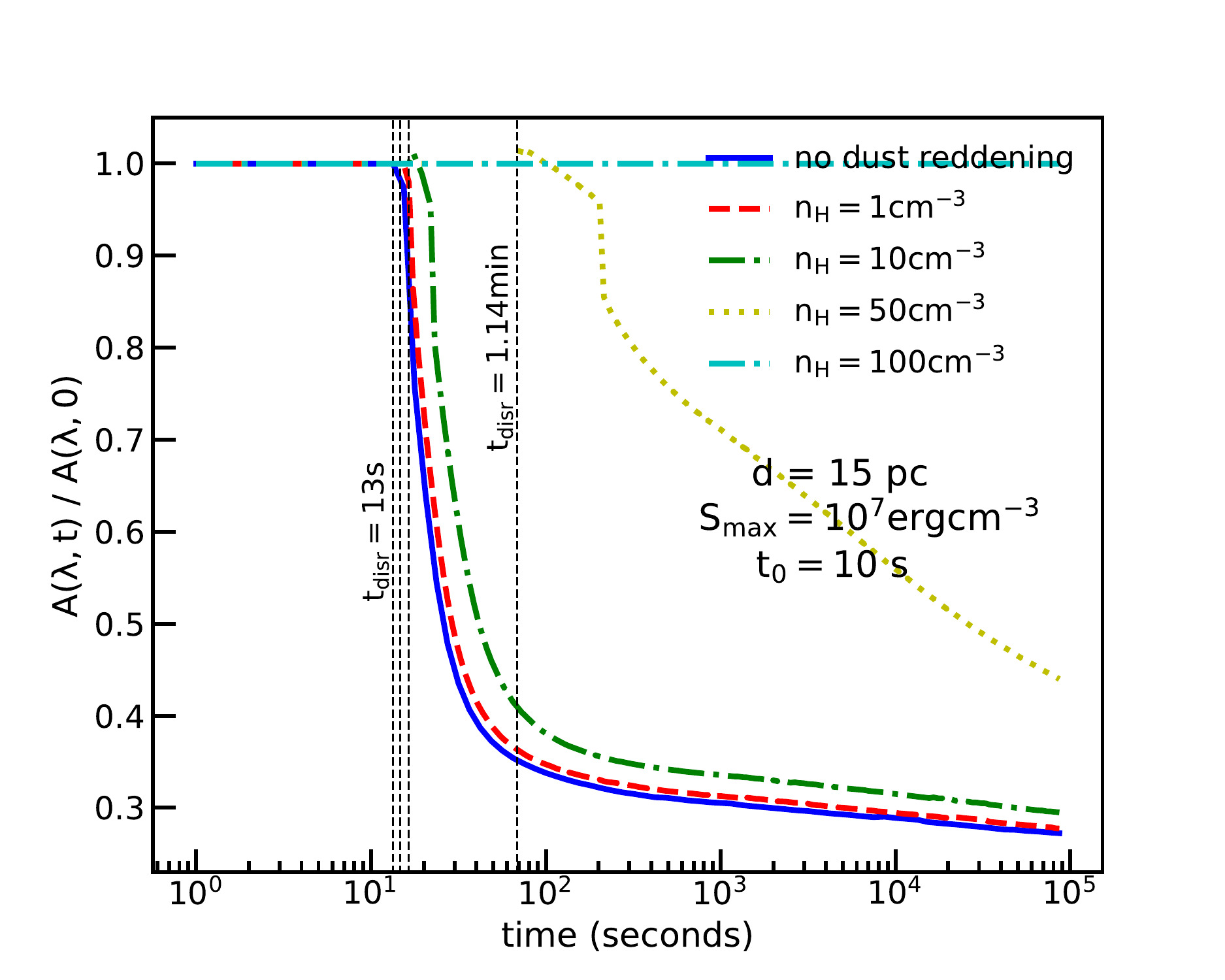}
\includegraphics[width=0.5\textwidth]{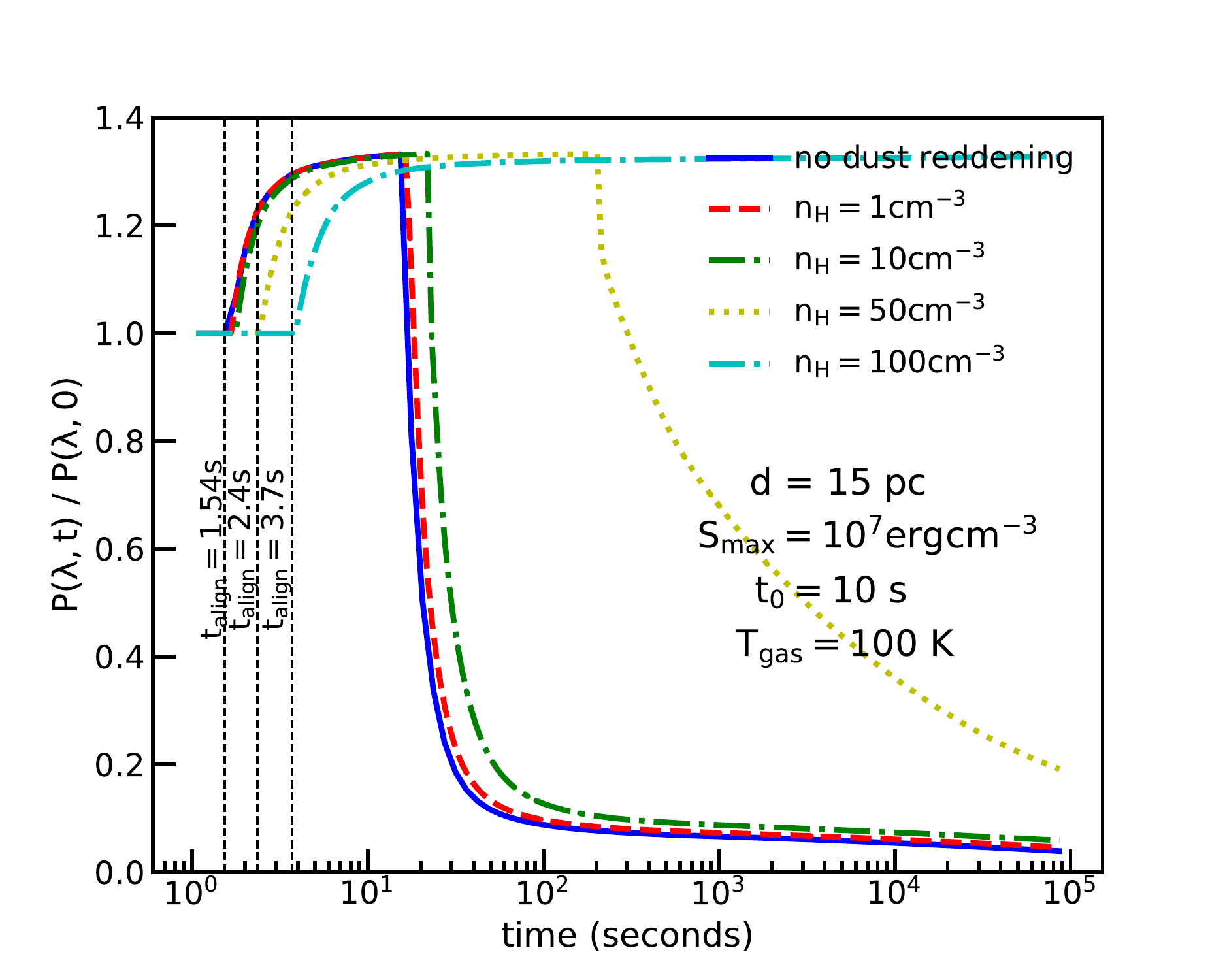}
\caption{Time-variation of the optical extinction $A_{V}$ and the optical polarization degree $P_{V}$ given by grains considered in Figure \ref{fig:disrGRB_nH}, assuming $S_{\rm max}=10^{7}\erg\cm^{-3}$ and $T_{\rm gas}=100$ K.}
\label{fig:Rv_GRB}
\end{figure}

As shown in Figure \ref{fig:adisr_time} (lower panel), the active region of RATD after one day is determined by $d_{\rm max,RATD} \sim$ 40 pc for $S_{\rm max}=10^{7}\erg\cm^{-3}$ and the peak time $t_{0}=10\s$. It corresponds to the radiation energy density $u_{\rm rad,min}=L_{\rm bol} e^{-\rm \tau}/(4 \pi c d^{2}) \sim 1.84\times 10^{-3}\erg\cm^{-3}$. Therefore, we use this value as the lower limit of the radiation strength $U$ that RATD can disrupt the maximum grain size of $0.25 \mum$. Besides, we assume that all slices have the same value of the effective optical depth $\tau$. Then, the new value of $d_{\rm max,RATD}$ from GRB afterglows in the presence of the dust reddening can be estimated by:
\begin{equation*}
d_{\rm max,RATD} = 10 {\rm pc} + N\Delta d,
\end{equation*}
where $N$ is the number of slice as determined by:
\bea 
u_{\rm rad,min} = 1.84 \times 10^{-3} \equiv \frac{L_{\rm bol} e^{-N \tau}}{4 \pi c (10 + N\Delta d)^{2}}.
\ena
 
The radiation energy density is reduced significantly when the gas column density $N_{\rm H}$ increases, reducing the active region of RATD. For instance, with $\Delta d=0.2$ pc, we find that $N \sim 105$ corresponds with $d_{\rm max,RATD}=30$ pc for $n_{\rm H}=10\cm^{-3}$, but $N$ will reduce to 35 that gives $d_{\rm max,RATD}=17$ pc for $n_{\rm H}=100\cm^{-3}$.

Figure \ref{fig:disrGRB_nH} (upper panel) shows the grain disruption size as a function of cloud distances with and without radiation attenuation after one day, assuming $S_{\rm max}=10^{7}\erg\cm^{-3}$ and $t_{0}=10$ s and different gas density $n_{\rm H}$. One can see that the active region of RATD decrease continuously from 40 pc to 35 pc and then to 15 pc when the gas density increases from $n_{\rm H} \ll 1$ (no disruption case) to $n_{\rm H}=1$, $10\cm^{-3}$ and $100\cm^{-3}$, respectively.

Figure \ref{fig:disrGRB_nH} (lower panel) shows the time-variation of the grain disruption size in the presence of dust attenuation in a cloud at 15 pc, assuming the same parameters as in the upper panel. The attenuation of radiation field by intervening dust causes the delay of grain disruption, but the effect is only significant for $n_{\H}\gtrsim 50\cm^{-3}$. 

Figure \ref{fig:Rv_GRB} shows the time variability of the optical extinction (upper panel) and optical polarization (lower panel), assuming a cloud at $d=15$ pc. The effect of radiation attenuation is marginal for low density of $n_{\H}\lesssim 10\cm^{-3}$ and becomes significant for $n_{\H}\gtrsim 50\cm^{-3}$.
 
\subsection{Effect of multiple clouds toward individual GRBs}

Our modeling results in Figure \ref{fig:Aext_time} show that the extinction curve becomes much steeper over time due to grain disruption. The parameter $R_{V}$ decreases with time accordingly (Figure \ref{fig:RV_Smax}). The small $R_{V}$ values can reproduce the steep far-UV rise extinction curves observed toward individual GRB afterglows (\citealt{Schady12}). However, our present results are obtained for a single-cloud model. In realistic situations, there may be more than one cloud along the line of sight toward a GRB afterglow. The effect of multiple clouds would not change the disruption time because it is only determined by the first cloud. However, it will change the amplitude of the variation in dust extinction and polarization.

Let $N_{\H}$ be the total hydrogen column density along a line of sight toward a GRB afterglow. Let $f_{\H}=N_{H}^{D}/N_{\H}$ with $N_{\H}^{D}$ being the hydrogen column density of the active region of RATD. The total extinction is given by: 
\bea
A(\lambda)&=&A(\lambda)^{D} + A(\lambda)^{ND}\nonumber\\
&=&N_{\H}^{D}\left(\frac{A(\lambda)}{N_{\H}}\right)_{D}+N_{\H}^{ND}\left(\frac{A(\lambda)}{N_{\H}}\right)_{ND}\nonumber\\
&=&N_{\H}\left[f_{\H}\left(\frac{A(\lambda)}{N_{\H}}\right)_{D}+(1-f_{\H})\left(\frac{A(\lambda)}{N_{\H}}\right)_{ND}\right],~~~~\label{eq:Atotal}
\ena
where $D$ and $ND$ stand for disruption and no-disruption region. This corresponds to:
\bea
\frac{A(\lambda)}{N_{\H}}=\left[f_{\H}\left(\frac{A(\lambda)}{N_{\H}}\right)_{D}+(1-f_{\H})\left(\frac{A(\lambda)}{N_{\H}}\right)_{ND}\right].
\ena

Since $(A(\lambda)/N_{\H})_{D,ND}$ only depends on dust content of the cloud, the observed total extinction per H and $R_{V}$ are determined by the parameter $f_{\H}$, i.e., the amount of dust in the closest cloud. Therefore, the observed value $R_{V}$ would be larger than predicted by a single cloud model in Figure \ref{fig:RV_Smax} for $f_{\H}<1$. 

One can obtain a similar relationship for dust polarization as follows:
\bea
\frac{P(\lambda)}{N_{\H}}=\left[f_{\H}\left(\frac{P(\lambda)}{N_{\H}}\right)_{D}+(1-f_{\H})\left(\frac{P(\lambda)}{N_{\H}}\right)_{ND}\right].~~~\label{eq:Ptotal}
\ena

Using detailed modeling of the extinction and polarization curves with observational data, we can constrain the distribution of matter along the line of sight toward GRB afterglows. This would shed light on the progenitors of GRBs.
 
\subsection{Origins of dark GRBs and microwave emission}
Extinction by intervening dust grains is a popular explanation for "dark" optical GRBs. In light of our study, we predict that some optical GRBs may be dark in the beginning but will become visible due to the decrease of optical/NIR extinction as a result of RATD. The time-variation monitoring of optical GRB afterglows would be useful to test this scenario. Moreover, we expect that intervening dust clouds should be far away from "dark" GRBs such that intense GRB afterglows cannot disrupt a considerable amount of dust via RATD (e.g., $d>40$ pc).

If GRBs are indeed located in a dusty star-forming region, then, within 40 pc from GRBs, the environment is likely dominated by very small grains (VSGs) due to the RATD effect. Such tiny grains would produce significant microwave emission between 10-100 GHz via spinning dust mechanism (\citealt{Drain98}; \citealt{Hoang10}; \citealt{2011ApJ...741...87H} \citealt{Hoang16}). Therefore, radio and microwave observations beyond a timescale of days would be useful to test RATD, shedding light on the origin of dark GRBs. An unsuccessful detection of spinning dust emission toward dark GRBs implies that dust clouds are very far from the source.

\section{Summary}\label{sec:sum}
We studied the rotational disruption of dust grains in the local environments of GRB afterglows using the Radiative Torque Disruption (RATD) mechanism and model extinction and polarization of GRB afterglows. Our main findings are summarized as follows:

\begin{enumerate}

\item For an optically thin medium, we show that large dust grains can be disrupted into smaller ones within one day up to 40 pc due to RATD. While thermal sublimation and Coulomb explosions only occur during the prompt phase of 10 s, RATD can disrupt grains by GRB afterglows at $t>10$ s.

\item We calculate the time-varying dust extinction of GRB afterglows in the presence of RATD. We find that the optical-NIR extinction decreases, whereas the UV and FUV extinction increases gradually until a day after the burst due to the enhancement of small grains by RATD. It causes the time-variability of color excess $E(B-V)$.

\item We model the polarization of GRB afterglows due to differential extinction by aligned grains. We show that the polarization first increases with time due to enhanced alignment by strong radiation fields and continues to change slowly when the grain disruption begins.

\item We compare our theoretical predictions with observational properties of GRB afterglows. We find that our predictions are in general supported by observations, including SMC-like extinction curves and low values of $R_{V}$ of GRB afterglows. Grain disruption by RATD can partly reproduces the optical re-brightening of GRB afterglows at late times.

\item Rotational disruption of large grains by GRB afterglows increases the abundance of very small grains in the local environment around GRBs. We suggest observing microwave emission from spinning dust toward GRB afterglows as new way to test RATD and the origin of dark GRBs.

\end{enumerate}

\acknowledgments
We thank E. Troja for discussions during the early stage of this work. This research was supported by the National Research Foundation of Korea (NRF) grants funded by the Korea government (MSIT) through the Basic Science Research Program (2017R1D1A1B03035359) and Mid-career Research Program (2019R1A2C1087045).

%--------------adding references-----------------------------------
\bibliographystyle{/Users/thiemhoang/Dropbox/Papers2/apj}
% or other styles: mcbride,plain, abbrv, acm, alpha, apalike, apj
\bibliography{/Users/thiemhoang/Dropbox/Papers2/cites_paperApJ,/Users/thiemhoang/Dropbox/Papers2/cites_Books}

\begin{thebibliography}{36}
\expandafter\ifx\csname natexlab\endcsname\relax\def\natexlab#1{#1}\fi

\bibitem[Abbas et al.(2004)]{2004ApJ...614..781A} Abbas, M.~M., Craven, P.~D., Spann, J.~F., et al.\ 2004, \apj, 614, 781

\bibitem[Andersson et al.(2015)]{Ander15} Andersson, B.-G., Lazarian, A., \& Vaillancourt, J.~E.\ 2015, \araa, 53, 501 

\bibitem[{Barth {et~al.}(2003)Barth, Sari, Cohen, Goodrich, Price, Fox, Bloom, Soderberg, \& Kulkarni}]{Barth03}
Barth, A.~J., Sari, R., Cohen, M.~H., {et~al.} 2003, \apj, 584, L47

\bibitem[{Berger {et~al.}(2003)}]{Berger03}
Berger, E., Kulkarni, S. R., Pooley, G., et al. 2003, Nature, 426, 154

\bibitem[Bolmer et al.(2018)]{2018A&A...609A..62B} Bolmer, J., Greiner, J., Kr{\"u}hler, T., et al.\ 2018, \aa, 609, A62

\bibitem[{Burke \& Silk(1974)}]{Burke74}
Burke, J. R., \& Silk, J. 1974, \apj, 190, 1
 
\bibitem[{Chiar {et~al.}(2006)Chiar, Adamson, Whittet, Chrysostomou, Hough,
  Kerr, Mason, Roche, \& Wright}]{Chiar06}
Chiar, J.~E., Adamson, A.~J., Whittet, D. C.~B., {et~al.} 2006, \apj, 651, 268

\bibitem[Covino et al.(2003a)]{Covino03a} Covino, S., Malesani, D., Tavecchio, F., et al.\ 2003a, \aa, 404, L5

\bibitem[Covino et al.(2003b)]{Covino03b} Covino, S., Malesani, D., Ghisellini, G., et al.\ 2003b, \aa, 400, L9

\bibitem[{Covino {et~al.}(2003)}]{Covi03}
Covino, S., Ghisellini, G, Malesani, D. et al, \gcn, 2167, 1C

\bibitem[{Covino \& Gotz(2016)}]{Covino16}
Covino, S., \& Gotz, D. 2016, \aa, 29, 205

\bibitem[Covino \& Gotz(2016)]{Covino:2016vo} Covino, S., \& Gotz, D.\ 2016, Astronomical and Astrophysical Transactions, 29, 205

\bibitem[de Ugarte Postigo et al.(2018)]{ugarte18} de Ugarte Postigo, A., Th{\"o}ne, C.~C., Bensch, K., et al.\ 2018, \aa, 620, A190

\bibitem[Dolginov, \& Mitrofanov(1976)]{1976Ap&SS..43..291D} Dolginov, A.~Z., \& Mitrofanov, I.~G.\ 1976, \apss, 43, 291

\bibitem[{Draine (2000)}]{Drain20}
Draine, B.~T. 2000, \apj, 532, 273

\bibitem[{Draine \& Hao(2002)Draine, \&
  Hao}]{Drain02}
Draine, B.~T., \& Hao, L. 2002, \apj, 569, 780

\bibitem[{Draine \& Lazarian(1998)}]{Drain98}
Draine, B.~T., \& Lazarian, A. 1998, \apj, 508, 157

\bibitem[{Draine \& Li(2007)}]{Drain07}
Draine, B.~T., \& Li, A. 2007, \apj, 657, 810

\bibitem[Draine, \& Weingartner(1996)]{1996ApJ...470..551D} 
Draine, B.~T., \& Weingartner, J.~C.\ 1996, \apj, 470, 551

\bibitem[{Draine, \& Salpeter(1979)}]{Draine79}
Draine, B. T., \& Salpeter, E. E. 1979, \apj, 231, 77

\bibitem[{Fraija {et~al.}(2019)Fraija, Dichiara, do~E~S~Pedreira, Galvan-Gamez,
  Becerra, Duran, \& Zhang}]{Fraija19}
Fraija, N., Dichiara, S., do~E~S~Pedreira, A. C.~C., {et~al.} 2019, arXiv.org

\bibitem[{Fruchter {et~al.}(2001)Fruchter, Krolik, \&
  Rhoads}]{Fruchter01}
Fruchter, A., Krolik, J.~H., \& Rhoads, J.~E. 2001, \apj,
  563, 597

\bibitem[Gehrels et al.(2009)]{Gehrels:2009} Gehrels, N., Ramirez-Ruiz, E., \& Fox, D.~B.\ 2009, \araa, 47, 567

\bibitem[{Greiner {et~al.}(2013)}]{Greiner13}
Greiner, J., Krühler, T., Nardini, M., et al. 2013, \aa, 560, A70

\bibitem[{Giang {et~al.}(2019)Giang, Hoang, \& Tram}]{Giang19}
Giang, N.~C., Hoang, T., \& Tram, L.~N. 2019, arXiv:1906.11498, \apj, in press

\bibitem[Heintz et al.(2017)]{Heintz:2017} 
Heintz, K.~E., Fynbo, J.~P.~U., Jakobsson, P., et al.\ 2017, \aap, 601, A83

\bibitem[Herranen et al.(2019)]{Herranen:2019kj} Herranen, J., Lazarian, A., \& Hoang, T.\ 2019, \apj, 878, 96

\bibitem[{Hoang(2017)}]{Hoang17}
Hoang, T. 2017, \apj, 836, 13

\bibitem[{Hoang(2019)}]{Hoang19b}
Hoang, T. 2019, \apj, 876, 13

\bibitem[{Hoang {et~al.}(2010)Hoang, Draine, \& Lazarian}]{Hoang10}
Hoang, T., Draine, B.~T., \& Lazarian, A. 2010, \apj, 715, 1462

\bibitem[Hoang, \& Lazarian(2008)]{Hoang:2008} 
Hoang, T., \& Lazarian, A.\ 2008, \mnras, 388, 117

\bibitem[Hoang, \& Lazarian(2009a)]{2009ApJ...695.1457H} 
Hoang, T., \& Lazarian, A.\ 2009, \apj, 695, 1457

\bibitem[Hoang, \& Lazarian(2009b)]{Hoang:2009} 
Hoang, T., \& Lazarian, A.\ 2009, \apj, 697, 1316


\bibitem[{Hoang \& Lazarian(2014)}]{Hoang14}
Hoang, T., \& Lazarian, A. 2014, \mnras, 438, 680

\bibitem[{Hoang \& Lazarian(2016)}]{Hoang16}
Hoang, T., \& Lazarian, A. 2016, \apj, 831, 159

\bibitem[{Hoang {et~al.}(2015{\natexlab{a}})Hoang, Lazarian, \&
  Andersson}]{Hoang15a}
Hoang, T., Lazarian, A., \& Andersson, B.-G. 2015{\natexlab{a}}, \mnras, 448,
  1178
  
 \bibitem[Hoang et al.(2011)]{2011ApJ...741...87H} Hoang, T., Lazarian, A., \& Draine, B.~T.\ 2011, \apj, 741, 87

\bibitem[Hoang et al.(2014)]{2014ApJ...790....6H} 
Hoang, T., Lazarian, A., \& Martin, P.~G.\ 2014, \apj, 790, 6
 
\bibitem[{Hoang {et~al.}(2015{\natexlab{b}})Hoang, Lazarian, \&
  Schlickeiser}]{Hoang15b}
Hoang, T., Lazarian, A., \& Schlickeiser, R. 2015{\natexlab{b}}, \apj, 806, 255

\bibitem[{Hoang {et~al.}(2013)Hoang, Lazarian, \& Martin}]{Hoang13}
Hoang, T., Lazarian, A., \& Martin, P.~G. 2013, \apj, 779, 152

\bibitem[Hoang et al.(2019)]{Hoang19} Hoang, T., Tram, L.~N., Lee, H., \& Ahn, S.-H.\ 2019, Nature Astronomy, 3, 766

\bibitem[{Hoang {et~al.}(2016)Hoang, Vinh, \& Quynh~Lan}]{HoangVinh16}
Hoang, T., Vinh, N.~A., \& Quynh~Lan, N. 2016, \apj, 824, 18

\bibitem[{Izzo {et~al.}(2019)Izzo, de~Ugarte~Postigo, Maeda, Th{\"o}ne, Kann,
  Della~Valle, Sagues~Carracedo, Micha{\l}owski, Schady, Schmidl, Selsing,
  Starling, Suzuki, Bensch, Bolmer, Campana, Cano, Covino, Fynbo, Hartmann,
  Heintz, Hjorth, Japelj, Kami{\'n}ski, Kaper, Kouveliotou, Kru{\.z}y{\'n}ski,
  Kwiatkowski, Leloudas, Levan, Malesani, Micha{\l}owski, Piranomonte,
  Pugliese, Rossi, S’́anchez-Ram’́irez, Schulze, Steeghs, Tanvir,
  Ulaczyk, Vergani, \& Wiersema}]{Izzo19}
Izzo, L., de~Ugarte~Postigo, A., Maeda, K., {et~al.} 2019, Nature, 565, 324

%\bibitem[{Jordana {et~al.}(2019)}]{Jordana19}
%Jordana, M., Mundell, C., Kobayashi, S., {et~al.}, 2019, arXiv:1911.08499

\bibitem[{Kann {et~al.}(2018)Kann, Schady, Olivares~E, Klose, Rossi, Perley,
  Zhang, Kr{\"u}hler, Greiner, Nicuesa~Guelbenzu, Elliott, Knust, Cano, Filgas,
  Pian, Mazzali, Fynbo, Leloudas, Afonso, Delvaux, Graham, Rau, Schmidl,
  Schulze, Tanga, Updike, \& Varela}]{Kann18}
Kann, D.~A., Schady, P., Olivares~E, F., {et~al.} 2018, \aa, 617, A122

\bibitem[Klotz et al.(2005)]{Klotz2005} Klotz, A., Bo{\"e}r, M., Atteia, J.~L., et al.\ 2005, \aap, 439, L35

\bibitem[{Kr{\"u}hler {et~al.}(2011)Kr{\"u}hler, Greiner, Schady, Savaglio,
  Afonso, Clemens, Elliott, Filgas, Gruber, Kann, Klose, K{\"u}pc{\"u}-Yolda{\c
  s}, McBreen, Olivares, Pierini, Rau, Rossi, Nardini, Nicuesa~Guelbenzu,
  Sudilovsky, \& Updike}]{Kruhler11}
Kr{\"u}hler, T., Greiner, J., Schady, P., {et~al.} 2011, \aa, 534, A108
  
\bibitem[Lazarian, \& Hoang(2007)]{2007MNRAS.378..910L} 
Lazarian, A., \& Hoang, T.\ 2007, \mnras, 378, 910
  
\bibitem[{{Lazarian} {et~al.}(2015){Lazarian}, {Andersson}, \&
  {Hoang}}]{Laza15}
{Lazarian}, A., {Andersson}, B.-G., \& {Hoang}, T. 2015, {Grain alignment: Role
  of radiative torques and paramagnetic relaxation}, ed. L.~{Kolokolova},
  J.~{Hough}, \& A.-C. {Levasseur-Regourd}, 81

\bibitem[{Laskar {et~al.}(2019)Laskar, Alexander, Gill, Granot, Berger,
  Mundell, Barniol-Duran, Bolmer, Duffell, van Eerten, Fong, Kobayashi,
  Margutti, \& Schady}]{Laskar19}
Laskar, T., Alexander, K.~D., Gill, R., {et~al.} 2019, arXiv.org,
  arXiv:1904.07261

\bibitem[{Loeb \& Perna(1998)}]{Loeb98}
Loeb, A., \& Perna, R. 1998, \apj, 495, 597

\bibitem[Mathis(1986)]{1986ApJ...308..281M} Mathis, J.~S.\ 1986, \apj, 308, 281

\bibitem[{Mathis {et~al.}(1977)Mathis, Rumpl, \& Nordsieck}]{Mathis77}
Mathis, J.~S., Rumpl, W., \& Nordsieck, K.~H. 1977, \apj, 217, 425

\bibitem[{{Mathis} {et~al.}(1983)}]{Mathis83}
{Mathis}, J. S., {Mezger}, P. G., \& {Panagia}, N. 1983, \aa ,500, 259

\bibitem[{Melandri {et~al.}(2017)Melandri, Covino, Zaninoni, Campana, Bolmer,
  Cobb, Gorosabel, Kim, Kuin, Kuroda, Malesani, Mundell, Nappo, Sbarufatti,
  Smith, Steele, Topinka, Trotter, Virgili, Bernardini, D'Avanzo, D'Elia,
  Fugazza, Ghirlanda, Gomboc, Greiner, Guidorzi, Haislip, Hanayama, Hanlon, Im,
  Ivarsen, Japelj, Jel{\'\i}nek, Kawai, Kobayashi, Kopac, LaCluyz{\'e},
  Martin-Carrillo, Murphy, Reichart, Salvaterra, Salafia, Tagliaferri, \&
  Vergani}]{Melandri17}
Melandri, A., Covino, S., Zaninoni, E., {et~al.} 2017, \aa, 607, A29

\bibitem[{Meszaros \& Rees(1997)}]{Meszaros97}
Meszaros, P., \& Rees, M.~J. 1997, \apj v.476, 476, 232

\bibitem[{Morgan {et~al.}(2014)Morgan, Perley, Cenko, Bloom, Cucchiara,
  Richards, Filippenko, Haislip, LaCluyze, Corsi, Melandri, Cobb, Gomboc,
  Horesh, James, Li, Mundell, Reichart, \& Steele}]{Morgan14}
Morgan, A.~N., Perley, D.~A., Cenko, S.~B., {et~al.} 2014, \mnras, 440, 1810

%\bibitem[{Mundell {et~al.}(2013)Mundell, Kopac, Arnold, Steele, Gomboc,
%  Kobayashi, Harrison, Smith, Guidorzi, Virgili, Melandri, \&
%  Japelj}]{Mundell13}
%Mundell, C.~G., Kopac, D., Arnold, D.~M., {et~al.} 2013, Nature, 504, 119

\bibitem[{Nardini {el~al.}(2011)}]{Nardi11}
Nardini, M., Greiner, J., Krühler, T., et al. 2011, \aa, 531, A39 

\bibitem[{Nardini {et~al.}(2014)}]{Nardi14}
Nardini, M., Elliott, J., Filgas, R., {et~al.}, 2014, \aa, 562, A29


\bibitem[{Paczy{\'n}ski(1998)}]{Paczy98}
Paczy{\'n}ski, B. 1998, \apj, 494, L45

\bibitem[{Perley {et~al.}(2010)}]{Perley10}
Perley, D. A., Bloom, J. S., Klein, C. R., et al. 2010, \mnras, 406, 2473

\bibitem[{Perna \& Lazzati(2002)}]{Perna02}
Perna, R., \& Lazzati, D. 2002, \apj, 580, 261

\bibitem[{Perna {et~al.}(2003)Perna, Lazzati, \& Fiore}]{Perna03}
Perna, R., Lazzati, D., \& Fiore, F. 2003, \apj, 585, 775

\bibitem[{Rol {et~al.}(2000)}]{Rol20}
Rol, E., Wijers, R.A.M.J, Vreeswijk, E. et al, \apj, 544, 707

\bibitem[{Schady(2017)}]{Schady17}
Schady, P. 2017, Royal Society Open Science, 4, 170304

\bibitem[{Schady {et~al.}(2010)}]{Schady10}
Schady, P., Page, M. J., Oates, S. R., et al. 2010, \mnras, 401, 2773 

\bibitem[{Schady {et~al.}(2012)Schady, Dwelly, Page, Kr{\"u}hler, Greiner,
  Oates, de~Pasquale, Nardini, Nardini, Roming, Roming, Rossi, Rossi, Still, \&
  Still}]{Schady12}
Schady, P., Dwelly, T., Page, M.~J., {et~al.} 2012, \aa, 537, 15

\bibitem[{Stratta {et~al.}(2013)Stratta, Gendre, Atteia, Boer, Coward,
  de~Pasquale, Howell, Klotz, Oates, \& Piro}]{Stratta13}
Stratta, G., Gendre, B., Atteia, J.~L., {et~al.} 2013, \apj, 779, 66

\bibitem[{Waxman \& Draine(2000)}]{Waxman20}
Waxman, E., \& Draine, B.~T. 2000, \apj, 537, 796

\bibitem[{Weingartner \& Draine(2001)}]{Wein01}
Weingartner, J.~C., \& Draine, B.~T. 2001, \apj, 548, 296

\bibitem[Weingartner et al.(2006)]{Weingartner:2006} 
Weingartner, J.~C., Draine, B.~T., \& Barr, D.~K.\ 2006, \apj, 645, 1188

\bibitem[Zafar et al.(2019)]{Zafar:2019} Zafar, T., Heintz, K.~E., Karakas, A., et al.\ 2019, \mnras, 490, 2599
\bibitem[Zafar et al.(2018)]{Zafar:2018} Zafar, T., Watson, D., M{\o}ller, P., et al.\ 2018, \mnras, 479, 1542
\end{thebibliography}
%\bibliography{ms.bbl}

\end{document}